%% file: sample-sigconf-isca23.tex
\documentclass[sigconf]{acmart}

\AtBeginDocument{%
  \providecommand\BibTeX{{%
    \normalfont B\kern-0.5em{\scshape i\kern-0.25em b}\kern-0.8em\TeX}}}

\copyrightyear{2023} 
\acmYear{2023} 
\setcopyright{rightsretained} 
\acmConference[ISCA '23]{Proceedings of the 50th Annual International Symposium on Computer Architecture}{June 17--21, 2023}{Orlando, FL, USA}
\acmBooktitle{Proceedings of the 50th Annual International Symposium on Computer Architecture (ISCA '23), June 17--21, 2023, Orlando, FL, USA}
\acmDOI{10.1145/3579371.3589047}
\acmISBN{979-8-4007-0095-8/23/06}

\include{commands}

\begin{document}
\title{OneQ: A Compilation Framework for Photonic One-Way Quantum Computation} 
 \author{Hezi Zhang}
 \email{hezi@cs.ucsb.edu}
 \affiliation{
     \institution{University of California}
     \city{Santa Barbara}
     \country{USA}
 }
  \author{Anbang Wu}
 \email{anbang@cs.ucsb.edu}
 \affiliation{
     \institution{University of California}
     \city{Santa Barbara}
     \country{USA}
 }
 \author{Yuke Wang}
 \email{yuke_wang@cs.ucsb.edu}
 \affiliation{
     \institution{University of California}
     \city{Santa Barbara}
     \country{USA}
 }
 \author{Gushu Li}
 \email{gushuli@ece.ucsb.edu}
 \affiliation{
     \institution{University of California}
     \city{Santa Barbara}
     \country{USA}
 }
 \author{Hassan Shapourian}
 \email{hshapour@cisco.com}
 \affiliation{
     \institution{Cisco Quantum Lab}
     \city{San Jose}
     \country{USA}
 }
 \author{Alireza Shabani}
 \email{ashabani@cisco.com}
 \affiliation{
     \institution{Cisco Quantum Lab}
     \city{Sant Monica}
     \country{USA}
 }
 \author{Yufei Ding}
 \email{yufeiding@cs.ucsb.edu}
 \affiliation{
     \institution{University of California}
     \city{Santa Barbara}
     \country{USA}
 }
\renewcommand{\shortauthors}{Hezi Zhang, et al.}

\input{01_abstract}



\maketitle

\input{02_introduction}

\input{03_background_title}
\input{03_background_hardware}
\input{03_background_software}

\input{04_motivation}

\input{05_algorithm_design}
\input{07_evaluation_corrected}

\input{08_conclusion}
\input{09_acknoledgement}

\bibliographystyle{unsrt}
\bibliography{references}

\end{document}

%% file: commands.tex
\usepackage{amsmath,amsthm, mathtools,amsfonts}
\usepackage{titlesec}
\DeclareUnicodeCharacter{2009}{\,} 
\titlespacing*{\section}
{0pt}{\parsep}{\parsep}
\titlespacing*{\subsection}
{0pt}{\parsep}{\parsep}
\usepackage{xcolor}

\newcommand\ket[1]{\left|#1\right\rangle}

\usepackage{algorithmic}
\usepackage{graphicx}
\usepackage{textcomp}
\usepackage{xcolor}

\usepackage[T1]{fontenc}
\usepackage[utf8]{inputenc}

\def\BibTeX{{\rm B\kern-.05em{\sc i\kern-.025em b}\kern-.08em
    T\kern-.1667em\lower.7ex\hbox{E}\kern-.125emX}}

\newcommand\yuke[1]{\textcolor{blue}{\textbf{[Yuke: #1]}}}

\newcommand\fix[1]{#1}

\usepackage[normalem]{ulem}

\usepackage{xcolor}
\usepackage{color}
\definecolor{codegreen}{rgb}{0,0.6,0}
\definecolor{codegray}{rgb}{0.5,0.5,0.5}
\definecolor{codepurple}{rgb}{0.58,0,0.82}
\definecolor{backcolour}{rgb}{0.95,0.95,0.92}
\definecolor{textblue}{rgb}{.2,.2,.7}
\definecolor{textred}{rgb}{0.54,0,0}
\definecolor{textgreen}{rgb}{0,0.43,0}
\definecolor{codered}{rgb}{201,72,12}
\usepackage[T1]{fontenc}
\usepackage[scaled=0.85]{beramono} 
\usepackage{listings}
\usepackage{multirow}
\usepackage{multicol}
\usepackage{url}
\usepackage{subfig}
\usepackage[ruled,vlined,linesnumbered]{algorithm2e}
\usepackage{multirow}
\usepackage{tikz}
\usepackage{xcolor}
\usepackage{ctable} 

\newtheorem{lemma}{Lemma}
\newcommand{\frameworkname}{OneQ}

\newcommand{\archName}{extendable space-time coupling graph}

\newcommand{\qsixteenGrid}{7x7}
\newcommand{\qsixteenRSG}{16x16}

\newcommand{\qtwentyfiveGrid}{9x9}
\newcommand{\qtwentyfiveRSG}{21x21}

%% file: 01_abstract.tex
\begin{abstract}

In this paper, we propose \frameworkname, the first optimizing compilation framework for one-way quantum computation towards realistic photonic quantum architectures. 
Unlike previous compilation efforts for solid-state qubit technologies, our innovative framework addresses a unique set of challenges in photonic quantum computing.
Specifically, this includes the dynamic generation of qubits over time, the need to perform all computation through measurements instead of relying on 1-qubit and 2-qubit gates, and the fact that photons are instantaneously destroyed after measurements.
As pioneers in this field, we demonstrate the vast optimization potential of photonic  one-way quantum computing, showcasing the remarkable ability of \frameworkname\ to reduce computing resource requirements by orders of magnitude.

\end{abstract}

%% file: 02_introduction.tex
\section{Introduction}
\label{sect:intro}

Quantum computing (QC) is promising in providing significant speedup for many problems, such as large-number
factorization \cite{shor1999polynomial}, unordered
database search \cite{grover1996fast} and simulation in quantum chemistry~\cite{quantum_chemistry_1,quantum_chemistry_2}.
Tremendous progress has been made in hardware technologies for quantum computing, including superconducting~\cite{devoret2013superconducting}, ion trap~\cite{bruzewicz2019trapped}, neutral atoms~\cite{saffman2016quantum}, and photonics~\cite{photonics_review}, in both research lab settings and industrial production levels.
With no clear winner among current platforms, it is expected that multiple technologies will continue to evolve together to meet diverse mission needs and application requirements.

In recent years, there has been a surge of interest in photonic quantum computing~\cite{takeda2019toward,pant2019percolation}, mainly due to the unique advantages of photonic qubits~\cite{OBrien2009,Bogdanov:17}, such as great scalability, long coherence time and easy integration with quantum networks.
 This interest is further fueled by the breakthroughs in photonic technologies~\cite{larsen2021deterministic, wang2020integrated} and the successful demonstration of quantum supremacy on photonic quantum computers~\cite{Madsen2022,zhong2020quantum,PhysRevLett.127.180502}.
Moreover, PsiQuantum~\cite{bartolucci2021fusion, bombin2021interleaving, bartolucci2021switch} recently announced the world’s first manufacturing milestone for integrated quantum photonic chips, with quantum photonic and electronic chips being manufactured using advanced semiconductor tools. 

However, photonic quantum computing also presents a unique set of features that cannot be directly handled by conventional programming and compiler frameworks~\cite{Qiskit,sivarajah2020t} for solid-state qubits. 
First, unlike the permanent qubits in solid-state systems~\cite{devoret2013superconducting,bruzewicz2019trapped,saffman2016quantum}, photonic qubits are flying qubits at the speed of light, and are generated dynamically over time.
%
Second, in contrast to the solid-state platforms which entangle qubits by multi-qubit gates such as CNOT, entanglement between photonic qubits are more natively performed by projective measurements onto entangled states, a process referred to as \emph{fusion} (e.g.,  simultaneous measurements in $XX$- and $ZZ$-bases project the qubits onto Bell states). 
Third, measurements on photonic qubits, including single-qubit measurements and fusions, instantaneously destroy the photons, meaning that each photonic qubit can be measured or fused at most once and cannot be reinitialized.

At programming level, measurement-based quantum computing (MBQC)~\cite{mbqc2009} has been suggested as a native programming paradigm for photonic quantum computing, because it does not require long-lasting qubits or quantum gates, but only one-time measurements on qubits.
Unlike the circuit model which initializes qubits to a product state of all $\ket{0}$s and performs algorithms by a sequence of gates that gradually creates entanglements, MBQC initializes qubits to an entangled state and performs algorithms by a pattern of single-qubit projective measurements (called \emph{measurement pattern}) that gradually consumes the entanglements.
During the computation, each qubit is measured only once and is then free to detach from the computing system or even disappear. This is why MBQC is also called ``one-way'' quantum computing (1WQC). 
\fix{In addition to MBQC, another popular programming paradigm for photonic quantum computing~\cite{Xanadu, Madsen2022} is the special-purpose model for Gaussian boson sampling~\cite{hamilton2017gaussian}, but it is not a universal computing model like MBQC and thus has restricted applications.}

Yet few attempts have been made on the compilation side. 
While previous work~\cite{mbqc2009,raussendorf2003measurement, briegel2009measurement,van2006universal} has shown that the utilization of a lattice-like entangled state, i.e., a \emph{cluster state}, can enable a basic interpreter from the circuit model to MBQC by joining the measurement patterns of individual gates 
on it, this approach is not efficient for scaling up on hardware. 
Practical photonic hardware scales up by first generating small entangled states (called \emph{resource states}) and then combining them into larger entangled states through fusions~\cite{GHZcluster}.
Attempting to generate a large cluster state on such hardware would result in an unnecessarily significant fusion overhead, because a large amount of entanglements in the cluster state is actually redundant for the computation, but only to keep the lattice geometry.
Instead of generating all entanglements of the cluster states, it is more efficient to only generate the entanglements required by the computation. Theses required entanglements can be represented by a non-lattice \emph{graph state}, i.e. a generalized cluster state with arbitrary graph geometry.
However, as the geometry of the graph state is irregular and varies depending on the specific program to be executed, it is not yet known how to efficiently map the irregular graph geometry onto an MBQC machine while complying with the hardware constraints.


In this paper, we investigate how to optimize the quantum program compilation for a photonic MBQC machine.
As the first step, we abstract a \emph{\archName} to formulate the hardware model. 
This coupling graph represents the hardware resources (e.g., resource state generators (RSG)) and constraints (e.g., available photon routing) in a way that suits the subsequent compilation designs.
With this abstraction, we identify three key challenges in compiling graph states onto the hardware model. 
First, resource states are generated dynamically in layers over time by limited number of RSGs and thus cannot be available simultaneously.
Second, there is no direct correspondence between a graph state node and a photonic qubit due to the lack of high-degree qubits in the resource states.
Third, there is a mismatch between the irregular geometry of the graph state and the regular structure of fusion supports represented by the coupling graph. 
Additionally, for resource states of small size, there is a further constraint that non-planar graphs cannot be accommodated by resource states generated in the same layer since the number of possible fusions on each resource state is too limited.

\begin{figure}
    \centering
    \vspace{5pt}
    \includegraphics[width=0.47\textwidth]{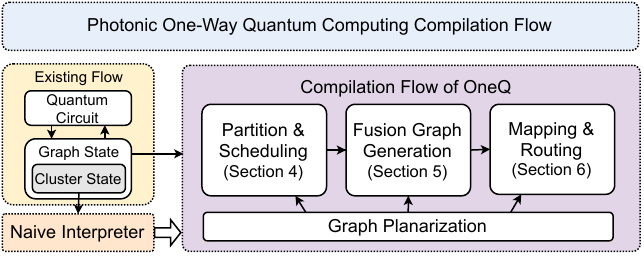}
    \caption{Framework Overview.}
    \label{fig:designoverview}
\end{figure}

To this end, we propose an optimizing compilation framework consisting of three sequential stages (Figure~\ref{fig:designoverview}), with each stage handled by a key module of the framework.
The \textbf{Graph Partition and Scheduling} module (Section~\ref{sect:module3}) partitions the graph state into subgraphs and schedules them onto resource states generated at different times.
It achieves a balance between reducing the wait time of photons and increasing the potential efficiency of mapping by analyzing the dependencies among operations and considering the overall geometry.
The \textbf{Fusion Graph Generation} module (Section~\ref{sect:module1}) synthesizes the scheduled subgraphs by fusing low-degree qubits in the resource states into high-degree nodes in the graph and connecting them appropriately. 
The produced fusion strategy is agnostic to the coupling constraints, and is represented by a \emph{fusion graph}.
%
The \textbf{Fusion Mapping and Routing} module (Section~\ref{sect:module2}) tackles the coupling constraints by finding an embedding layout of the irregular fusion graph into the regular coupling graph.
Specifically, it makes the fusion graph compatible with the coupling graph by extending and routing the edges of the fusion graph along the edges of the coupling graph.
By maximizing the compactness of the layout, we can reduce the number of required fusions, thus reducing the computation overhead and enhancing the overall fidelity.
Lastly, the additional constraint for small resource states can be addressed by a planarity enforcement for all modules.
%
%
With these modules and optimizations, our framework maximizes the utilization of computing resources and efficiently accommodates the irregularity and adaptivity of the graph states.

To summarize, our contributions in this paper are listed as below:
\begin{itemize}
    \item We abstract a \archName\ to formulate the hardware model of a realistic photonic MBQC architecture.
    
    \item We identify three critical compilation challenges in the gap between MBQC programs and realistic photonic MBQC hardware, as well as an additional constraint in the cases of small resource states.

    \item We propose an optimizing compilation framework to efficiently deploy quantum programs onto the hardware, which includes three key optimization modules to tackle the identified challenges, along with a planarity enforcement process to address the additional constraint.
    
    \item Our framework outperforms the basic MBQC interpreter by orders of magnitude in terms of execution time and resource consumption of the compiled program.
\end{itemize}

%% file: 03_background_title.tex
\vspace{0.3em}
\section{Background and Related Work}\label{sect:background}

%% file: 03_background_hardware.tex
\vspace{0.3em}
\subsection{MBQC on Photonic Platforms}
\fix{
Photonic qubits are an exceptional carrier of quantum information and are well-suited for the one-way model of quantum computing~\cite{con-rev-255, con-rev-256}. The feasibility of photonic one-way quantum computing has been shown by implementing various quantum algorithms on photonic qubits in cluster states~\cite{con-rev-209}. For example, Grover's algorithm was implemented on a 4-photon 4-qubit cluster state in \cite{mbqc-grover} with a fidelity of 0.9, Deutsch-Jozsa algorithm was realized on a six-qubit cluster state in \cite{mbqc-dj} with a fidelity of 0.96, 
and Simon's algorithm was performed on a five-qubit cluster state in \cite{mbqc-simon} for a period-finding problem.
Notably, rapid progress is being made toward scaling up the photonic platform with integrated waveguides and optical chips~\cite{con-rev-102,con-rev-103,con-rev-104,con-rev-217,con-rev-204, ferreira2022deterministic, con-rev-247}, and a fully reconfigurable optical processor has been demonstrated in \cite{con-rev-252}.
}

\begin{figure}[h!]
    \centering
    \includegraphics[width=0.8\linewidth]{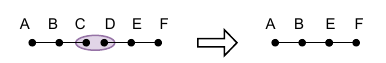}
    \caption{A fusion between graph states $ABC$ and $DEF$. After the fusion between $C$ and $D$, qubit $C$ and $D$ vanish while qubit $A,B,E$ and $F$ form a new graph state.}
    \label{fig:fusion_XZ_ZX}
\end{figure}

\fix{Practical photonic hardware is scaled up by generating small-size resource states (e.g., three-qubit~\cite{GHZcluster}, four-qubit graph states~\cite{psiquantum-fbqc}) and cascading them via fusion\fix{~\cite{con-rev-316}}.}
Fusion is a native operation in linear optics that projects a two-qubit state to an entangled state through joint measurements on the two qubits. For example, joint measurements in XX- and ZZ-basis projects two-qubit states onto Bell states. The joint measurements we use in this paper are in XZ- and ZX-basis, which merge two graph states to another graph state. As shown in Figure~\ref{fig:fusion_XZ_ZX}, the 3-qubit graph states $ABC$ and $DEF$ can be merged into a 4-qubit graph state $ABEF$ via a fusion on qubits $C$ and $D$. The cost of each fusion is losing the two measured photons (qubit $C$ and $D$ in Fig.~\ref{fig:fusion_XZ_ZX}). In general, the fusion between an $m$-qubit graph state and an n-qubit graph state generates an $(m+n-2)$-qubit graph state, which is larger than the original graph states when $m,n>2$.

Despite facilitating scalability, fusions come with an overhead of losing two qubits each time and are the lowest-fidelity operations in the computing process. 
As a result, reducing the number of fusions is critical for reducing the required computing resources and enhancing the overall fidelity. 
Thus we consider it as one of essential optimization goals in our compilation framework, which is analogous to the goal of minimizing the
number of CNOT gates in compilers for the circuit model.

%% file: 03_background_software.tex

\vspace{0.3em}

\subsection{Programming Paradigm of MBQC}

\vspace{0.3em}
\subsubsection{MBQC Basics}
MBQC is a universal but conceptually distinct computational model from the circuit model, as computation in MBQC is driven by single-qubit projective measurements instead of 1-qubit and 2-qubit gates. The initial state used  as the computing resource in MBQC can be a \emph{graph state}, which is an entangled state of qubits located on a graph $G=(V,E)$, formally defined as the eigenstates of operator\[s=X_i \bigotimes_{j\in n_i} Z_j, \quad\forall i\in V\] where $n_i$ is the set of neighboring qubits of $i\in V$. A graph state can be generated by preparing all qubits in $\ket{+}$ and applying a CZ gate on each pair of qubits connected by an edge $e\in E$.
Given this graph state, computation can be driven by a pattern of Z-measurements and \emph{equatorial measurements} $E(\alpha)$, i.e., measurements on the X-Y plane of Bloch sphere at an angle $\alpha$.
The graph $G$ and the measurement basis of each qubit together form a \emph{measurement pattern}.

MBQC has a classical-quantum hybrid nature in that it relies on a classical feed-forward to address the non-determinism of quantum measurement.
Specifically, the state of unmeasured qubits are subjected to Pauli X- or Z-corrections according to the outcomes of the measured qubits.
In practice, this can be implemented equivalently by adaptively adjusting the measurement bases of the remaining qubits, such that all corrections can be postponed to the end of the computation. 
As a result, these \emph{adaptive measurements} induce dependencies between measurements.
In particular, an $X^s Z^t$-correction with $s,t\in\{0,1\}$ can be postponed by adjusting its measurement angle from $\alpha$ to $(-1)^s\alpha+t\pi$, i.e., $E(\alpha)X^s Z^t = E((-1)^s\alpha+t\pi)$. %
The dependencies in $s$ and $t$ are called X- and Z-dependency respectively.

\begin{figure}[h!]
    \centering
    \includegraphics[width=0.96\linewidth]{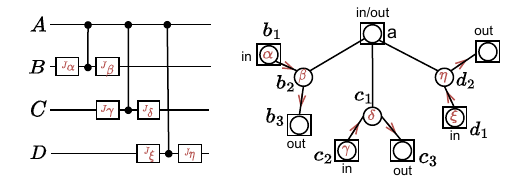}\\
    \hspace{0pt}(a)\hspace{100pt}(b)
    \caption{Translation from a circuit (a) to a graph state (b), with the measurement bases labeled on qubits. Qubits in (b) with `in', `out' or `in/out' are input or output qubits. Arrows on the edges indicate the dependencies between measurements.}
    \label{fig:circuit_translation}
\end{figure}

There is a straightforward translation~\cite{translation} from a circuit in the universal gate set $\{J(\alpha), \mathrm{CZ}\}$ into a measurement pattern, because a measurement $E(\alpha)$ yielding an outcome $m\in\{0,1\}$ is equivalent to a $J(\alpha)$ operator followed by a Z-measurement $M_z$ yielding the same value, i.e., $E(\alpha)\ket{\psi}\rightarrow m$~corresponds to $M_z J(\alpha)\ket{\psi}\rightarrow m$,
\begin{align*}
J(\alpha)=
\begin{bmatrix}
1 & \mathrm{e}^{\mathrm{i} \alpha}\\
1 & -\mathrm{e}^{\mathrm{i} \alpha}
\end{bmatrix}   
\end{align*}
For example, Figure~\ref{fig:circuit_translation} shows the translation from ~\ref{fig:circuit_translation}(a) to ~\ref{fig:circuit_translation}(b), with the measurement angles of qubits and the roles of being input or output marked on each of them. The 3-degree node $a$ in ~\ref{fig:circuit_translation}(b) corresponds to the qubit $A$ with 3 CZ gates in ~\ref{fig:circuit_translation}(a), while nodes $b_1, b_2$ correspond to single-qubit gates on qubit B (similarly for qubit C and D).
Dependencies between measurements are represented by directed edges.
For example, $b_1\rightarrow b_2$ indicates that the measurement basis of $b_2$ may be adjusted according to the measurement outcome of $b_1$. This process can be rigorously described in ZX-calculus and optimized by available tools such as PyZX~\cite{kissinger2019pyzx}.

\vspace{0.3em}
\subsubsection{MBQC Interpreter}
While there has not been a compiler for practical MBQC machines, previous work on cluster states enables a basic interpreter from the circuit model to MBQC, which we adopt as the baseline approach in our evaluation.

\begin{figure}[h!]
    \centering
    \includegraphics[width=\linewidth]{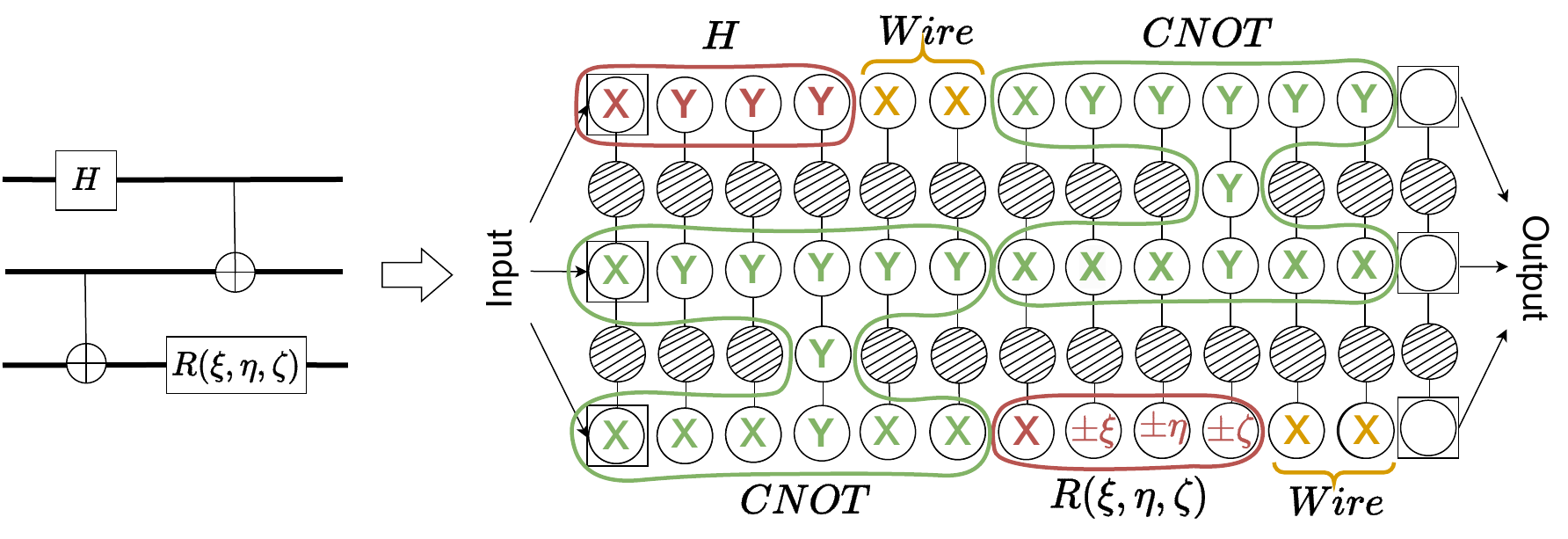}\\
    \hspace{-50pt}(a) \hspace{120pt} (b) 
    \caption{From a circuit (a) to an MBQC measurement pattern (b). Labels on the qubits stand for their measurement bases, with $X,Y$ standing for X-,Y-basis, and $\xi,\eta,\zeta$ standing for the angles of equatorial measurements. Shaded qubits are redundant qubits which can be removed by Z-measurements.}
    \label{fig:circuit_simulation}
\end{figure}

Cluster states are universal for MBQC as it allows the implementation of a universal gate set, such as single-qubit gates along with CNOT.
%
In particular, an arbitrary single-qubit gate can be implemented by a measurement pattern on a line of qubits in the cluster state, such as the red circles in Figure~\ref{fig:circuit_simulation}(b), while a CNOT gate can be implemented on two lines of qubits joined by a qubit in between, such as the green circles in Figure~\ref{fig:circuit_simulation}(b).
%
The $X,Y$ labels stand for X-, Y-basis, while $\xi,\eta,\zeta$ stand for the angles of general equatorial measurements. Measurements in bases except X, Y and Z are adaptive measurements and can thus induce dependency. These adaptive measurements only arise in the measurement patterns of non-Clifford gates, meaning that all Clifford gates can be executed simultaneously on the cluster state even if they are dependent in the circuit model.
For more details about measurement patterns of different gates, please refer to \cite{mbqc2009}.

A basic interpreter from the circuit model to MBQC can be constructed by simply joining the inputs and outputs of measurement patterns of those gates according to their positions in the circuit, as shown in Figure~\ref{fig:circuit_simulation}. Those measurement patterns can be aligned up by inserting identity gates, implemented by two adjacent X-measurements (or three adjacent Y-measurements), as shown by the yellow `wires' in Figure~\ref{fig:circuit_simulation}(b). The qubits not used by measurement patterns of any gates are redundant to the computation, and can be removed by measurements in Z-basis (shaded qubits in Figure~\ref{fig:circuit_simulation}(b)).
Similar to the circuit model compilation~\cite{sabre}, CNOT gates on non-neighboring qubit strips can be implemented by inserting measurement patterns of SWAP gates whenever necessary. In this way, we can implement any quantum circuit in an MBQC manner.
This interpreter greatly simplifies the mapping from qubits and gates in the circuit to the measurements on the cluster state. However, this approach is inefficient in the utilization of computing resources. Specifically, it makes many qubits redundant for computation and consumes their entanglements trivially by Z-measurements, without any regard for the huge overhead of generating those redundant entanglements on a practical MBQC machine.

%% file: 04_motivation.tex
\section{Problem Formulation}

\input{04_hardware_model}

\input{04_oneq_overview}

%% file: 04_hardware_model.tex
\subsection{Hardware Abstraction}
\label{sect:hardware_model}
Photonic hardware for one-way quantum computing involves three key stages: generation, routing, and measurement. In the first stage, an array of  \emph{resource state generators (RSG)} periodically produces copies of resource states, with the period referred to as a \emph{clock cycle}. In the second stage, routers channel the resource states to different measurement locations using \emph{spatial} and \emph{temporal routing}. Spatial routing allows fusions between resource states generated by different RSGs, while temporal routing allows fusions between resource states generated at different times using \emph{delay lines}, which can delay the arrivals of photons at the measurement devices for some clock cycles. In the third stage, computation is performed by measuring photons in a predetermined pattern, with the outcomes fed into a classical computer to determine the bases of subsequent measurements.

It can be seen that photonic hardware has distinct features that set it apart from solid-state hardware. Specifically, photonic qubits are continuously generated over time and consumed by measurements. The use of identical resource states simplifies hardware manufacture and enhances efficiency, but they are typically in a small size~\cite{GHZcluster,psiquantum-fbqc} and the number of resources states generated in each clock cycle is limited by the number of RSGs. Moreover, photonic hardware features two types of connections: spatial and temporal, both of which support fusions but are subject to different constraints. 
For example, spatial routing may only occur between resource states generated by neighboring RSGs, while temporal routing may only occur between resource states generated by the same RSG within a limited delay time. 

\begin{figure}[h!]
    \centering
    \includegraphics[width=\linewidth]{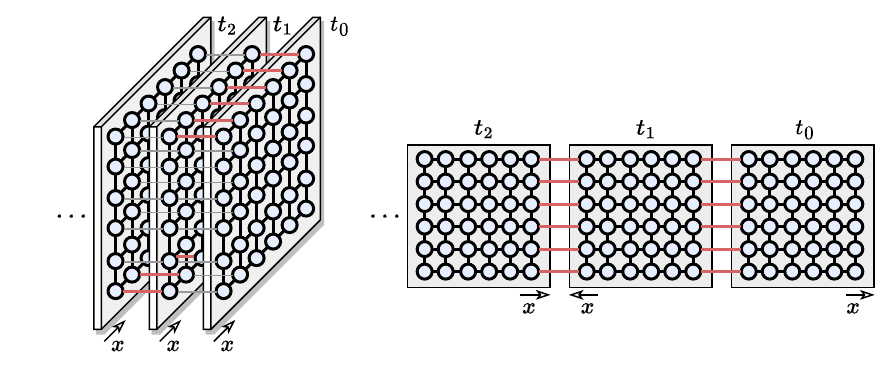}\\
    (a)\hspace{100pt}(b)
    \caption{3D structure of hardware connectivity (a) and its \archName~(b). The physical layer $t_1$ in (b) is flipped in the horizontal direction so that the physical layers $t_0,t_1$ and $t_2$ and the their temporal connections in red form an extended physical layer. To keep the figure clear, other temporal connections are omitted in (b).}

    \label{fig:archoverview}
\end{figure}

To capture these features, we formulate a hardware model abstraction called the \textit{\archName}, as illustrated in Figure~\ref{fig:archoverview}. Each node in the coupling graph represents a resource state generated by the RSG in the corresponding location, with resource states in different layers generated at different times. Each edge of the coupling graph represents a support for possible fusion operations. There are three levels of coupling, 
which we describe in detail below.

\textbf{Coupling in the resource state} The first level of coupling comes from entanglements of qubits in the same \emph{resource state}. These entanglements are generated directly by the RSGs and do not require fusion operations. As a result, this level of coupling is implicit in Figure~\ref{fig:archoverview}.
 

\textbf{Coupling within physical layers}
The second level of coupling comes from the resource states generated within the same clock cycle, which together form a layer of qubits referred to as a \emph{physical layer}.
As shown in Figure~\ref{fig:archoverview}, the nodes in the physical layer at $t_0$ (or $t_1,t_2$) are connected into a 2D lattice structure.
Such connections within a clock cycle represent the spatial connections, which allows a resource state to conduct a fusion with resource states generated at the same time by its neighboring RSGs.


\textbf{Coupling across physical layers}
The third level of coupling is that a resource state can make a fusion with another resource state generated in a different clock cycle by the same RSG, up to the limit of the delay time, as shown by the red and grey vertical lines in Figure~\ref{fig:archoverview}(a).
Note that we should avoid using too long-range temporal connections because photons suffer more from the loss error as they stay longer in the delay lines. 

In the coupling graph, the physical layers are extendable by combining consecutive physical layers and keeping their temporal connections at boundary. For example, in Figure~\ref{fig:archoverview}, the three physical layers (\ref{fig:archoverview}(a)) generated at time $t_0,t_1$ and $t_2$ can form an \emph{extended physical layer} by keeping their spatial connections and the temporal connections in red (\ref{fig:archoverview}(b)), with physical layer $t_1$ being flipped in the $x$ direction. Temporal connections other than the red ones are ignored in this extension and thus are not explicitly shown in Figure~\ref{fig:archoverview}(b). This extendability makes the area of physical layers flexible, and can allow mapping of more global structures onto the physical layers.

Despite the similarity to the quantum hardware coupling graph \cite{sabre} prevailing used for today’s superconducting and ion trap hardware architectures, we emphasize on the differences between them as following.

1) Different levels of coupling are not independent, and actually their interplay could create some additional constraints. For example, suppose the resource states generated by RSGs are 3-qubit resource states, then although the fusion device allows each location on the physical layers to maximally connect to 4 spatial neighbors (up, down, left, right) and the corresponding locations on previous and future physical layers up to the limit of delay lines, only up to 3 of them can be activated.

2) The temporal connections are convertable to spatial connections in the extended physical layers. The 3D connectivity structure of the coupling graph can be flattened into a series of 2D extended physical layers along the time dimension, with each extended layer consisting of several consecutive physical layers. Within each extended physical layer, the temporal connections (red lines in Figure~\ref{fig:archoverview})  are treated in the same way as spatial connections.

%% file: 04_oneq_overview.tex
\subsection{Overview of \frameworkname}\label{sect:overview}
\begin{figure*}[t!]
    \centering
\includegraphics[width=0.95\linewidth]{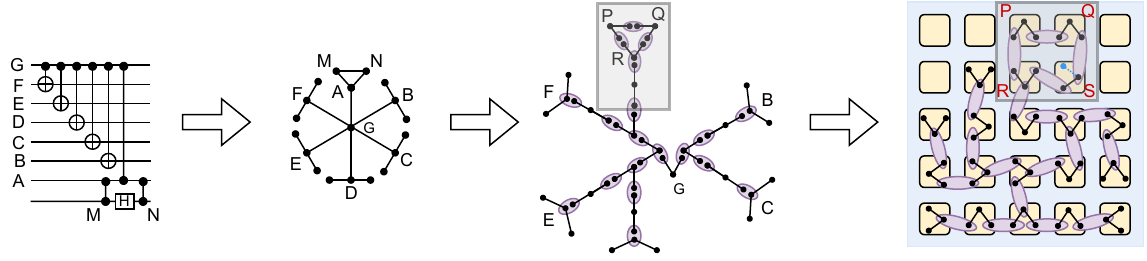}\\
    \hspace{-20pt}
    (a)\hspace{100pt}(b)\hspace{120pt}(c)\hspace{140pt}(d)
    \caption{Compilation flow. (a) Quantum circuit. (b) Graph state. (c) Fusion graph generation. (d) Fusion mapping and routing.
    }
    \label{fig:procedure}
\end{figure*}

As the first end-to-end compiler for MBQC, it is important to figure out a set of meaningful optimization goals. Our investigation on photonic one-way computing and realistic photonic quantum device development suggests the following two key optimization objectives.
    \textbf{1) Minimizing the physical depth}. The physical depth is defined as the number of physical layers consumed in the computation, which represents the time of executing the compiled program.
    \textbf{ 2) Reducing the number of fusions}. Fusions are often the most low-fidelity and costly operations incurred in such an MBQC photonic quantum computer.

In the rest of this subsection, we use a motivating example (shown in Figure~\ref{fig:procedure}) to give high-level ideas about the key challenges introduced in Section~\ref{sect:intro} and the insights of how they can be addressed by \frameworkname. We have attempted to maximize simplicity and ease of understanding, while possibly sacrificing
some rigor. We often omit the dependency arrows and explicit measurement bases when they are not important for our optimization techniques, and they can always be derived from the graph state \cite{translation}.

\textbf{Challenge 1: Dynamic Computing Resources}As resource states are generated dynamically in physical layers, a large number of qubits cannot be available simultaneously. In real-world quantum programs, graph states are typically too large (in terms of their number of nodes and edges) to be mapped onto a single or a few physical layers. Therefore, careful partition and scheduling schemes are needed to decompose large graph states into smaller components so that the graph state can be accommodated by the coordination of multiple physical layers. To minimize the delay time of photonic qubits and decrease their loss rates, dependent measurements can be scheduled according to their execution orders. However, this approach may compromise the compactness of graph state mapping onto the hardware by resulting in fragmented structures that lose the overall geometry of the graph state.
In \frameworkname, this challenge is addressed by an approach that concurrently considers both the dependencies between measurements and the overall program geometry, allowing for a more efficient and effective allocation of resources.
Section~\ref{sect:module3} gives detailed schemes of partition and scheduling.


\textbf{Challenge 2: Lack of High-Degree Qubits}
High-degree nodes are a common occurrence in graph states, as qubits performing many two-qubit gates with other qubits (e.g. node $G$ in Figure~\ref{fig:procedure}(a)) can result in high-degree qubit nodes in the corresponding graph state (e.g., node $G$ in Figure~\ref{fig:procedure}(b)). However, such high-degree qubits may not be available in the resource states due to the limited inter-connections among the photonic qubits they contain (e.g., qubits in the 3-qubit resource states in Figure~\ref{fig:procedure}(c) have a maximum degree of 2). This prevents a direct correspondence between the graph state nodes and the photonic qubits.
In \frameworkname, those high-degree qubit nodes are synthesized by a proper strategy of fusions between low-degree qubits in resource states (e.g., node $A$ to the shaded area in Figure~\ref{fig:procedure}(c), and node $G$ to the area around $G$ in the central part of Figure~\ref{fig:procedure}(c)). By matching with some basic fusion patterns, arbitrary high-degree input graph states (e.g., Figure~\ref{fig:procedure}(b)) can be synthesized to accommodate input adaptivity. Such a fusion strategy is agnostic to the routing constraints on the hardware, and aims to establish a viable fusion relation between resource states to synthesize the graph state eventually. The strategy can be abstracted to a \textit{fusion graph} that indicate necessary fusions at the level of resource state. 
Section~\ref{sect:module1} gives detailed schemes for the generation of synthesis strategies and their fusion graphs. 

\textbf{Challenge 3: Mapping and
Routing Constraints}
Due to the routing constraints, the irregular structure of fusions in Figure~\ref{fig:procedure}(b) may not be compatible with the hardware connectivity. In Figure~\ref{fig:procedure}(d), each yellow square represents a location of RGS, with the spatial routing allowing fusions between resource states from neighboring RGS's.
It can be seen that the triangle in Figure~\ref{fig:procedure}(c) consisting of resource states $P,Q$ and $R$ (circled part in Figure~\ref{fig:procedure}(c)) cannot be mapped to the orthogonal grid directly because $R$ can not be adjacent to both $P$ and $Q$ if $P$ and $Q$ are mapped to adjacent RSG locations. Such incompatibility exists widely in a general fusion graph, which prevents a direct mapping from the fusion graph to the underlying hardware topology. In \frameworkname\, this issue is resolved by extending some of the edges and winding them along the 2D lattice of individual or extended physical layers, which form paths of consecutive fusions. We refer to this procedure as fusion routing. For example, in Figure~\ref{fig:procedure}(d), $P,Q$ and $R$ can conduct fusions with each other with an auxiliary resource state $S$. In $S$, two qubits participate in the routing path $RSQ$, with the other qubit (in blue) removed by a Z-measurement. By maximizing the compactness of the mapping, we can minimize the overhead from extra fusions. Section~\ref{sect:module2} gives detailed searching schemes to map the edges onto orthogonal paths while minimizing the routing overhead.

\textbf{Additional Challenge: Non-Planar Graph States} The number of possible fusions on resource states is limited by the number of qubits they contain. Typically, for small resource states, each location in the coupling graph can be passed by at most one routing path. For example, in Figure~\ref{fig:procedure}(d), resource state $S$ is passed by the routing path of edge $QR$, which destroys two of its qubits. This leaves only one qubit remaining, which is insufficient for use in another routing path. Therefore, the layout of the graph state cannot have edges intersecting each other unless the resource states are large enough to be splitted into multiple 2-qubit cluster states (e.g., as large as 5-qubit linear cluster state or 6-qubit ring-like cluster state). In other words, only \emph{planar graphs}, i.e., graphs that can be embedded on a plane with no edges crossing each other, can be mapped to a single physical layer. In \frameworkname, this is addressed by enforcing a planarization process in all modules to prohibit the graph non-planarity.

%% file: 05_algorithm_design.tex
\input{05_scheduling_camera_ready}
\input{05_fusion_graph_generation_camera_ready}

\input{05_hezi_module2_camera_ready}

%% file: 05_scheduling_camera_ready.tex
\section{Graph Partition and Scheduling}
\label{sect:module3}
Scheduling of operations for MBQC differs from that for the circuit model in two ways. 
First, the dependency relations in MBQC can be quite different from those in the circuit model.
%
For example, dependencies between measurements occur only in the presence of non-Clifford gates, while all measurements corresponding to Clifford gates can be executed simultaneously.
Second, the number of graph state nodes that can be accommodated by each physical layer is flexible, which is hard to predict beforehand since it depends on the strategies of graph synthesis and routing.
Specifically, the synthesis of a node in the graph state may result in multiple nodes in the fusion graph, while the mapping of an edge in the fusion graph may require a routing path through multiple locations in the coupling graph.

To handle these distinct features, we partition the graph state into subgraphs in a way that respects the measurement dependency, and allow a dynamic scheduling for each subgraph to accommodate the flexibility in the resource demand. 
The graph partition enables the qubits measurable earlier to be scheduled onto earlier physical layers in a coarse grain, which reduces the wait time of qubits in the delay lines and enhances the program fidelity.
The dynamic scheduling schedules each partition of the graph state onto a flexible number of physical layers, with the physical layers allocated dynamically until the partition is all mapped.
This enables the search for a compact layout of the scheduled partition, which increases the efficiency of resource utilization and reduces the total number of physical layers required by the program.

The partition of the graph state respects the measurement dependency through an analysis of the executability order of the measurement operations. 
Due to the non-deterministic nature of quantum measurements, MBQC requires a classical feed-forward that induces dependencies between measurements.
The partial order between measurements can be described by a causal flow~\cite{translation}, which is determined by two types of dependencies.
Specifically, based on the measurement outcome of a particular qubit, the measurement angles of its neighboring qubits may be adjusted from $\alpha$ to $(-1)^s\alpha+t\pi$ where $s,t\in\{0,1\}$, with the dependency on $s$ called $X$-dependency and that on $t$ called $Z$-dependency.
Note that changing the angle from $\alpha$ to $\alpha+\pi$ only interchanges $\ket{\pm_\alpha}$ and $\ket{\mp_\alpha}$, which means the $Z$-dependency is equivalent to a re-interpretation of the measurement outcome of the affected qubit. Therefore the condition when a measurement becomes executable can be summarized as Lemma~\ref{lemma:executable_rule}, with qubits executable at each time forming a \emph{dependency layer}.

\begin{lemma}
\textit{A measurement on a qubit in the graph state is executable if all its $X$-dependent qubits are measured and all the $Z$-dependent qubits of all its $X$-dependent qubits are measured.}
\label{lemma:executable_rule}
\vspace{5pt}
\end{lemma}

This respect for measurement dependency is only in a coarse-grained manner. 
This means that a partition can contain multiple dependency layers, and dependency layers within the same partition do not have to be scheduled strictly according to their executability orders.
The reason for this is two-fold. 
First, the presence of delay lines can allow the delay of qubits' arrival at the measurement devices for some time, making an exact match between the executability order and the physical layer order unnecessary. 
Second, allowing for this mismatch enables a preservation of the overall geometry structure among nearby dependency layers, which can be used to increase the potential compactness of the graph layout in the mapping process.
Specifically, this coarse-grained manner can be achieved by first sorting the measurements according to their executability orders, and then forming each graph partition by including a number of consecutive dependency layers.

Partitioned subgraphs are then scheduled for fusion graph generation and layout search for the fusion graph in ascending order of their executability. The scheduling of these subgraphs is dynamic in two levels. First, for each partition, physical layers would be allocated dynamically for mapping and routing until all the nodes in the corresponding fusion graph are mapped to the hardware. Second, among partitions, there can be additional edges connecting nodes of fusion graphs for different partitions, which also need to be mapped to the hardware to form the whole graphs state. This can be achieved by dynamically allocating additional physical layers between the partitions, and routing on those physical layers until all cross-partition edges are mapped.

\textbf{Graph Planarization}
To further respect the planarity constraint for resource states of small size, the planarity of the subgraphs can be used as an additional condition for graph partitioning. Specifically, in the process of grouping consecutive dependency layers into partitions in ascending order of their executability, a planarity check can be performed along the way. New dependency layers will not be included if either the limit number of dependency layers is exceeded or the accumulated subgraph is no longer a planar graph. If a single dependency layer is already non-planar, it can be decomposed into subgraphs by repeatedly finding the maximal planar subgraph $G_i$ from its remaining graph $G$, where a maximal planar subgraph $G_i$ is defined as a subgraph such that the addition of any edge $e\in G-G_i$ breaks the planarity of $G_i$. This planarization can ensure in advance that each partition scheduled for fusion graph generation and layout searching is a planar graph, avoiding unnecessary conflicts when the graph is large and complex.

%% file: 05_fusion_graph_generation_camera_ready.tex
\vspace{0.3em}
\section{Fusion Graph Generation}
\label{sect:module1}


A graph state can be synthesized from small resource states via fusion by first synthesizing necessary components such as high-degree nodes and lines, then connecting them according to the graph geometry. In this section, we lay the foundation of synthesizing arbitrary graph states by defining a set of basic patterns. Then we demonstrate the process of graph synthesis from 3-qubit resource states and generalize it to more generic resource states.

\begin{figure}[h!]
    \centering
    \includegraphics[width=0.7\linewidth]{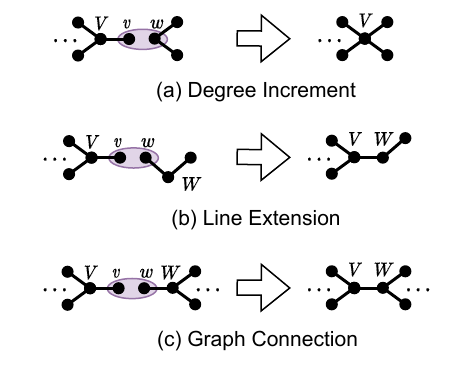}
    \caption{Basic fusion patterns.} 
    \label{fig:basic_fusion_patterns}
\end{figure}

Synthesis of arbitrary graph states can be realized by appropriate strategies based on some basic fusion patterns, as shown in Figure~\ref{fig:basic_fusion_patterns}. First, the degree of a node $V$ can be increased by a \emph{degree increment} pattern, which fuses a leaf node (degree $= 1$) connecting to $V$ with a node of degree $> 1$. For example, the degree of node $V$ in Figure~\ref{fig:basic_fusion_patterns}(a) is increased by 1 by fusing node $v$ with a 2-degree node $w$. Second, a line of nodes can be extended by a \emph{line extension} pattern, which fuses a leaf node of the line with that of another line. For example, the 2-node line $Vv$ in Figure~\ref{fig:basic_fusion_patterns}(b) is extended to a 3-node line by fusing the leaf node $v$ with the leaf node $w$. Third, two resource states can be connected while maintaining their structures (except the fused nodes) by a \emph{graph connection} pattern, which fuses two leaf nodes from each of them respectively. For example, the two 4-node graphs in Figure~\ref{fig:basic_fusion_patterns}(c) are connected by a fusion between node $v$ and $w$, with the nodes $V$ and $W$ connected by an edge in the formed graph state and the remaining structures of them left unchanged.

\begin{figure}[h!]
    \centering
    \includegraphics[width=0.65\linewidth]{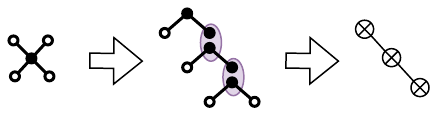}\\
    (a)\hspace{50pt}(b)\hspace{50pt}(c)
    \caption{Synthesize a high degree node in (a) by a basic fusion pattern in (b). Represent the fusion pattern by a fusion graph in (c), with each `$\otimes$' node representing a resource state, each edge representing a fusion operation.}
    \label{fig:linear_fusion}
\end{figure}

\begin{figure*}[h!]
    \centering
    \includegraphics[width=0.95\linewidth]{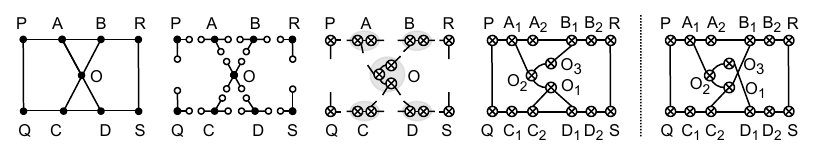}\\
    (a)\hspace{85pt}(b)\hspace{85pt}(c)\hspace{85pt}(d)\hspace{110pt}(e)
    \caption{Fusion graph generation. (a) A planar embedding of the graph state geometry. (b) Split the edges. (c) Replace each high-degree node with a piece of fusion graph, with each `$\otimes$' node representing a resource state. (d) Connect the pieces of fusion graphs while keeping the clockwise (counter-clockwise) edge orders in the planar embedding so as to obtain a planar fusion graph. (e) The planarity may be broken if not keeping the clockwise (counter-clockwise) edge orders.}
\label{fig:vertex_split}
\end{figure*}

With these basic patterns, an arbitrary graph state can be synthesized from resource states as small as 3-qubit resource states.
Specifically, high-degree nodes in the graph states can be synthesized by a repetitive application of the degree increment pattern until the required degree is satisfied, as shown in Figure~\ref{fig:linear_fusion}(a)(b).
The resulting fusion pattern can be represented by a \emph{fusion graph} in Figure~\ref{fig:linear_fusion}(c), with each node representing a resource state, and each edge representing a fusion operation.
Similarly, lines can be synthesized by a repetitive application of the line extension pattern until the required length is satisfied.
Then, the high degree nodes and lines can be connected via the graph connection pattern (Figure~\ref{fig:basic_fusion_patterns}(c)). We demonstrate it with an example of synthesizing a graph in Figure~\ref{fig:vertex_split}(a). In Figure~\ref{fig:vertex_split}(b), we break down the graph into high degree nodes $A,B,C,D,O$ and lines $PQ,RS$. Each high-degree node or each line is connected with some nodes in hollow dots, which represent its neighbors in Figure~\ref{fig:vertex_split}(a). Those nodes are referred to as \emph{virtual nodes}. Note that multiple virtual nodes are allowed to represent the same node. We synthesize the high-degree nodes and lines with the basic patterns Figure~\ref{fig:basic_fusion_patterns}(a)(b), pairing up the virtual nodes according to the original graph in Figure~\ref{fig:vertex_split}(a), and connect these components by applying the basic pattern Figure~\ref{fig:basic_fusion_patterns}(c) on the paired virtual nodes. \label{subsect:graph_synthesis}

For generic resource states, this synthesis process can be made applicable by a modification containing two steps.
First, nodes of the fusion pattern in Figure~\ref{fig:linear_fusion}(c) can be reduced by utilizing qubits of degree $> 2$.
%
%
In particular, for resource states with max degree $m$, an $n$-degree node can be synthesized by $n // m + 1$ resource states, where the extra $(n$ mod $m)$ qubits would be removed by Z-measurements.
Therefore, the fusion graph in Figure~\ref{fig:linear_fusion}(c) becomes a linear graph containing $n // m + 1$ nodes, instead of $n-1$ nodes in the case of 3-qubit resource states.
Second, edges in the fusion graph for 3-qubit GHZ states can be contracted by utilizing the connections between qubits within the resource states.
This can be achieved by a pattern matching process that traverses the graph, searching for lines or cycles that match the intrinsic geometry of the resource states.
%
During the synthesis process, qubits in resource states can be removed by Z-measurements when necessary. For example, a $n$-qubit ring-like resource state can be tailored to a ($n-1$)-qubit linear resource state by removing one qubit, if linear structures, rather than rings, need to be synthesized. 

\textbf{Planarity Preservation} The planarity condition imposed in the graph partition can be preserved in the process of fusion graph generation. In other words, if the original graph is planar, then the resulting fusion graph can also be planar.
This can be achieved by first finding a planar embedding of the original graph, i.e., a way of drawing it on the plane with no crossing edges, and then keeping the clockwise (counterclockwise) orders among the edges when connecting the synthesized high-degree nodes. The search for the planar embedding can be efficiently done by existing tools such as Networkx~\cite{networkx}.
For example, in the planar graph in Fig.~\ref{fig:vertex_split}(a), the 4-degree node $O$ has neighbors $A,B,D,C$ in the clockwise order. So we replace it with a fusion graph component consisting of 3 nodes (component $O$ in the grey circle at the center of Fig.~\ref{fig:vertex_split}(c)), and connect it with fusion graph components of $A,B,D,C$ in a clockwise manner (Fig.~\ref{fig:vertex_split}(d)). 
As a comparison, Fig.~\ref{fig:vertex_split}(e) shows a fusion graph that breaks planarity since the clockwise (counterclockwise) edge orders are not maintained.


%% file: 05_hezi_module2_camera_ready.tex
\section{Fusion Mapping and Routing}
\label{sect:module2}

Finding the most compact layout of a graph on a grid is known to be an NP-hard problem~\cite{compactness_NPhard}, and it is particularly challenging in our case due to two additional features of the grid. First, each layer of grid has limited area because each physical layer of resource states is generated by limited number of RSGs. Second, the space-time coupling in the hardware has a 3D nature, which makes a direct search of mapping inefficient. In particular,
each edge of a node can be mapped or routed toward not only different directions around the node but also any of the previous or subsequent layers up to the limit of delay time, which increases the search space and leads to longer search times.

\begin{figure}[h!]
    \centering
    \includegraphics[width=0.8\linewidth]{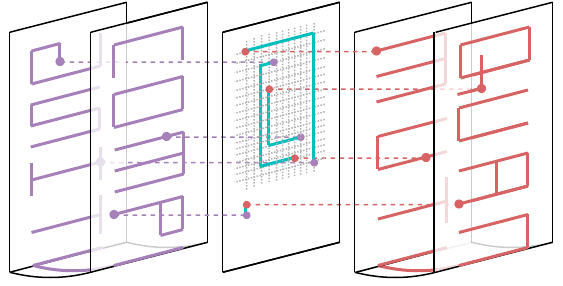}\\
    (a)\hspace{50pt}(b)\hspace{50pt}(c)
    \caption{Mappings of `QUANTUM' (red) and `COMPUTING' (purple) connected through a shuffling layer (b) in between. Red dots on (c) represent the incomplete nodes whose edges are not all mapped yet, with the other ends of them being the purple dots on (a). Corresponding nodes are connected by a shuffling on the intermediate layer (green lines in (b)).}
    \label{fig:shuffling}
\end{figure}

To address these challenges, we propose an algorithm consisting of an \emph{in-layer mapping} and an \emph{inter-layer shuffling}, which is applicable for grids of various geometry (triangular, orthogonal, etc.). 
It finds the 3D layout of the scheduled graph by finding a series of 2D compact layouts and connecting the them by a shuffling of nodes.
Each 2D layout is found by a boundary-aware heuristic search restricted to a physical layer or extended physical layer. 
The \emph{incomplete} nodes on each layer, whose edges are not all mapped due to the limited grid area are then connected by a node shuffling on additional physical layer between the 2D layouts.
For example, in Figure~\ref{fig:shuffling}, the graph `QUANTUM' (red) and `COMPUTING' (purple) are mapped onto two extended physical layers in \ref{fig:shuffling}(c) and \ref{fig:shuffling}(a), respectively. Red dots on \ref{fig:shuffling}(c) represent the incomplete nodes whose edges are not all mapped yet, with the other ends of those edges being the incomplete nodes represented by the purple dots on \ref{fig:shuffling}(a). These incomplete nodes are connected by a shuffling on the intermediate layer \ref{fig:shuffling}(b), with the green lines being the routing paths between corresponding nodes.

In the in-layer mapping, each edge is mapped onto the physical layer or extended physical layer either directly or via routing. 
The routing is triggered in two scenarios. 
First, it can map the edges between existing nodes in the layout, thus allowing the fusion between non-neighboring nodes and accommodating the irregularity in the structure of the scheduled graph. 
Second, it is used to reduce the blockage between mapped nodes and future nodes, thus increasing the number of edges that can be mapped to the layer and maximizing compactness of the layout.
A node is defined as \emph{blocked} if the number of its remaining unmapped edges $r$ exceeds the number of its adjacent available positions on the grid $s$, with the case of $s>0$ called \emph{partially blocked} and the case of $s=0$ called \emph{totally blocked}.
When a node is partially blocked by other mapped nodes, we can route the node to a position that has enough available degrees.
When the direct mapping of an edge result in a total blockage for some other nodes in the map, we can route the node to a broader area before the mapping, such that it leaves enough degrees for the other nodes.

\begin{figure}[h!]
    \newcommand{\fhe}{0.2}
        \includegraphics[height=\fhe\textwidth]{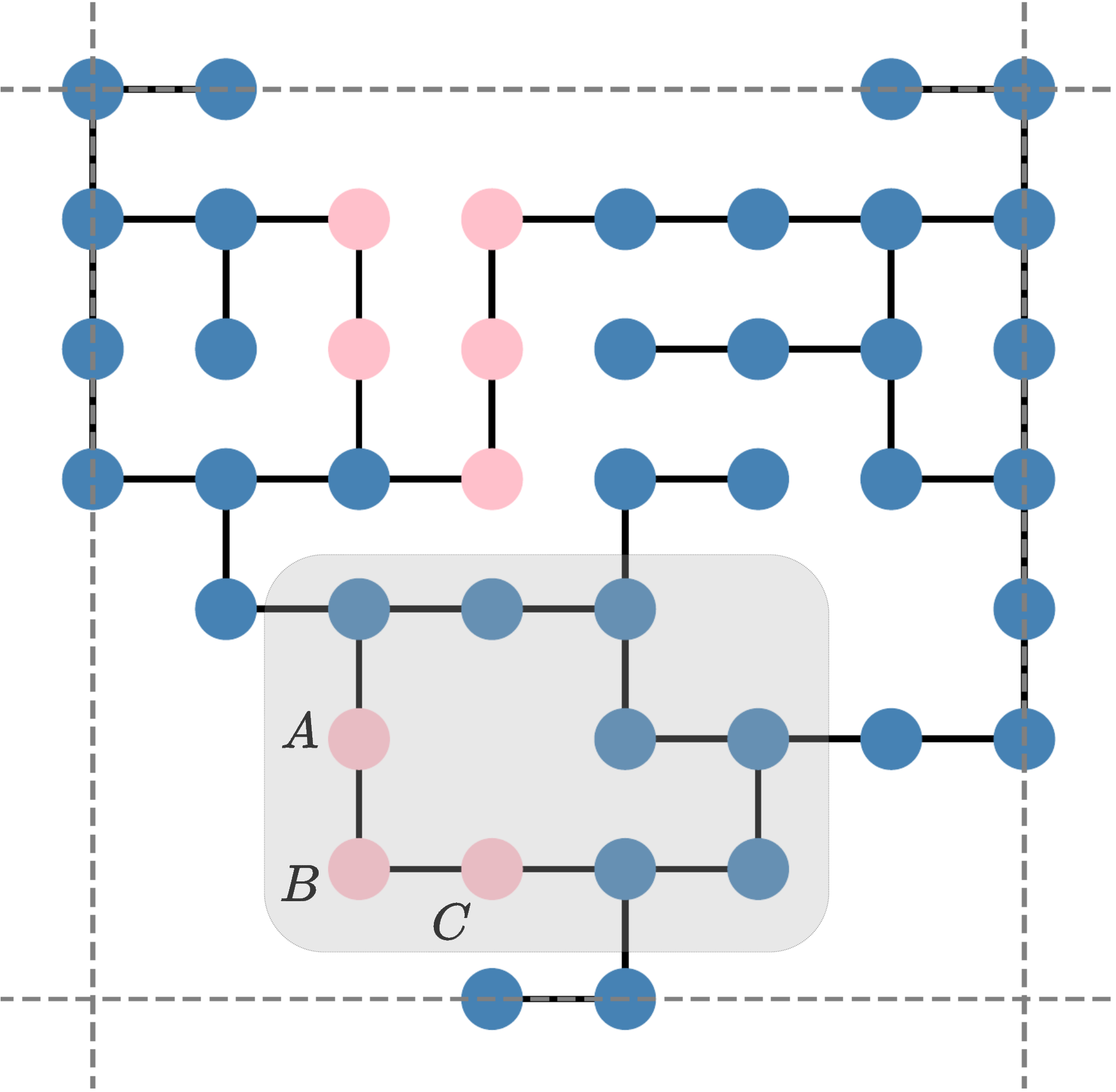}
        \includegraphics[height=\fhe\textwidth]{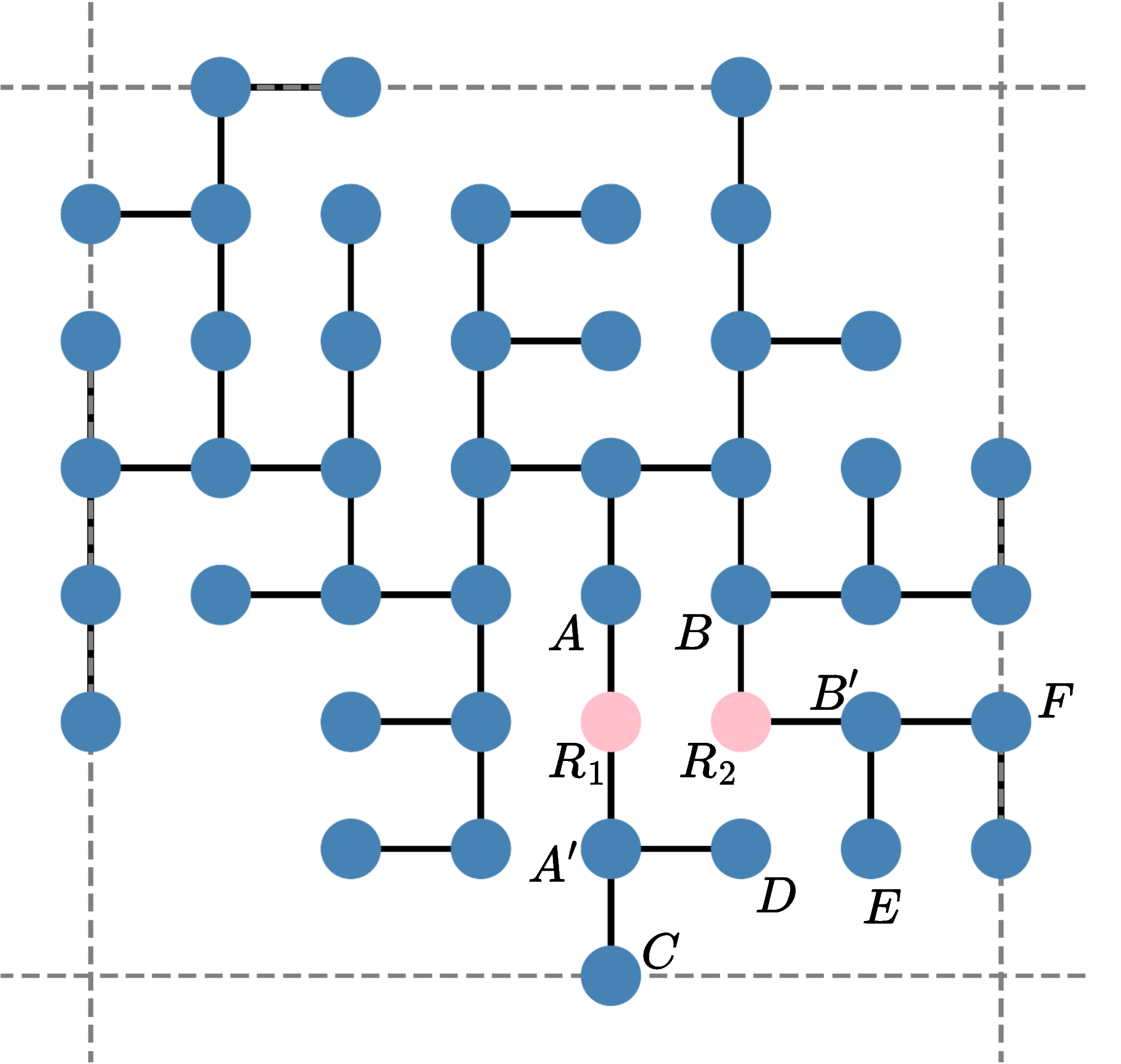} 
        \\
        \hphantom{}(a) \hspace{70pt} (b) \hspace{200pt}
        \caption{(a) Mapping of fusion graph for 8-qubit BV with secret string `11111111'. (b) Mapping of fusion graph for 3-qubit QFT. Blue and green dots correspond to the nodes `$\otimes$' existing in the fusion graph, with the green ones being the incomplete nodes whose edges are not all mapped yet. Pink dots represent the auxiliary resource states used for routing.
        }
        \label{fig:case_study}
\end{figure}

This can be more clearly demonstrated with two layout examples.
In Figure~\ref{fig:case_study}(a), the routing path $ABC$, represented by pink nodes, is used to connect the existing nodes and form the cyclic structure in the shaded area. In Figure~\ref{fig:case_study}(b), routing was triggered for twice. First, node $A'$ is routed from $R_1$ by one step to avoid the total blockage for node $B$, because if node $A'$ is directly mapped to the location $R_1$, one of its edges $A'C$ and $A'D$ would have to take location $R_2$, which is the only remaining location around $B$. Second, node $B'$ is routed by one step to address its partial blockage by $A'D$, because while node $B'$ can be directly mapped to the location $R_2$, the only one remaining location around it would be insufficient for mapping its two future edges $B'E$ and $B'F$. Note that in real cases the length of routing paths should be $\ge 2$.

The heuristic search algorithm can be stated as below. We traverse the edges in a cycle-prioritized breadth-first order, giving priority to edges involved in cycles over those in trees, because tree edges are more flexible for mapping. We choose to map each edge either directly or via routing according to the scenarios above. After each new edge is mapped, the layout is evaluated by the heuristic cost function
\begin{align*}
    H = \enspace&\mathrm{occupied\_area}\\
    &+ \#\mathrm{partially\_blocked\_nodes}\\
&+ \alpha\cdot\#\mathrm{totally\_blocked\_nodes}
\end{align*}
where $\mathrm{occupied\_area}$ refers to the area of the smallest rectangle that can enclose the mapped edges. $\alpha$ should be larger than 1 as a total blockage is more harmful. It can be typically set as the maximum degree of each physical layer. The layout with the lowest cost is then selected as the best candidate and used in the search for future edges.

In the inter-layer shuffling, incomplete nodes in different 2D layouts are connected via routing on the physical layers allocated between them. This shuffling can be achieved by first pairing up the incomplete nodes, sorting the node pairs according to their distances, and then finding the shortest routing paths to connect the node pairs in ascending order of the distances. Specifically, if a single physical layer is not enough to accommodate all the routing paths, more physical layers can be allocated dynamically until all the node pairs are connected. The inter-layer shuffling can be used to support the subgraph connection need arising in two circumstances. First, as the graph partition process partitions the whole graph state into subgraphs, the fusion graphs of different subgraphs need to be connected to form the whole graph state. Second, the mapping of each partition may generate multiple 2D layouts, as some nodes may be inevitably blocked due to the limited area of physical layers and the complexity of the graph geometry. While positions of incomplete nodes in the previous layers are taken into account in the mapping of subsequent layers to increase their proximity, those 2D layouts may still need to be connected by the inter-layer shuffling if not all conflicts can be resolved within the in-layer mapping.



\textbf{Planarity-Aware Search}
When the given fusion graph is planar, the heuristic search for in-layer mapping can be made planarity-aware to maximize the number of edges accommodated by each physical layer.
Specifically, we can first find a planar embedding of the fusion graph, and make 
the mapping of each edge follow the planar embedding, which can be achieved by forcing the mapped edge to leave enough degrees for its clockwise or counterclockwise neighboring edges connected to the same node. 
This planarity-awareness naturally prevents routing paths from crossing each other and reduces the demand for inter-layer shuffling.
This approach is more effective when the area of physical layers is large enough, because otherwise the edges would be blocked by the limited area anyway. 
This area can be increased by extension of physical layers, as shown in Figure~\ref{fig:archoverview}(b).

%% file: 07_evaluation_corrected.tex
\vspace{0.3em}
\section{Evaluation}\label{sect:eval}

\vspace{0.3em}
\subsection{Experiment Setup}\label{sect:expset}

\noindent\textbf{Metric} As discussed in Section~\ref{sect:overview}, the first metric we consider is the physical depth, i.e., the number of physical layers consumed in the execution of the program. 
The second metric is the number of fusion operations performed in the execution of the program, which is denoted as `\#~fusion'. A better compiler should compile a quantum program into a smaller depth which indicates a shorter execution time and less consumption of resources, and a smaller `\#~fusion' which indicates less overhead and being less error-prone. For comparison with the baseline, we define an improvement factor of physical depth as the ratio of the depth in the baseline to the depth in \frameworkname, and an improvement factor of `\#~fusion' as the ratio of the number of fusions consumed in the baseline to the that consumed in \frameworkname.

\noindent \textbf{Baseline} 
Our optimized baseline is based on the basic MBQC interpreter introduced in Section~\ref{sect:background} but with two preprocessing steps.
The first step is synthesizing a sufficiently large 3D cluster state by fusions from small resource states generated by the RSGs. The sufficient size of each 2D layer of the cluster state is referred to as \emph{cluster area}, while the number of RSGs required to form a cluster state of that size is referred to as \emph{physical area}. The second step is finding a layout for the circuit so that measurement patterns of all gates in the circuit can be joined in a way shown as Figure~\ref{fig:circuit_simulation}. 

For the first step, instead of relying on any specific synthesis strategy, we adopt a lower bound of the physical area, which indicates the least number of required resource states when ignoring the spatial constraints for fusion routing.
For the second step, to ensure an efficient implementation of a two-qubit gate with MBQC, it is necessary to place the strips of the involved qubits close to each other in the cluster state, similar to that in the circuit model with solid-state-qubit quantum computing. As a result, we utilize the qubit mapping and routing functions of Qiskit~\cite{Qiskit} to address the issues related to far-apart 2-qubit gates prior to mapping the circuit onto the cluster state.
Specifically, this is achieved by using a 2D coupling graph that shares the same structure with each 2D layer of the cluster state as the hardware coupling graph to Qiskit.

\noindent \textbf{Benchmark programs} We select Quantum Fourier transform (QFT), quantum approximate optimization algorithm (QAOA), Ripple-Carry Adder (RCA)~\cite{RCA} and Bernstein-Vazirani (BV) algorithm as our benchmarks, which include both building blocks of quantum programs (QFT and RCA) and application driven programs targeting at solving real-world problems (QAOA and BV).
For QAOA,
we choose the graph maxcut problem on randomly generated graphs. Specifically, the graphs are generated by randomly selecting half of its all possible edges.
For BV, we select the secret strings randomly, with approximately half of the digits being 0 and half being 1. In table~\ref{tab:benchmark}, we list the number of qubits and the number of gates  for each benchmark, as well as the cluster area and physical area required by the baseline approach. For each benchmark, our framework is evaluated on the same physical area with the baseline to make them more comparable.

\begin{table}[h]
    \centering
      \caption{Benchmark programs.}
      
    \resizebox{0.43\textwidth}{!}{
    \renewcommand*{\arraystretch}{1.4}
    \begin{tabular}{|p{1.8cm}|p{1.0cm}|p{1.5cm}|p{1.8cm}|}  \hline
    Name & \#qubit & cluster area & physical area  \\ \hline

  \multirow{3}{1.8cm}{Quantum Fourier Transform (QFT)} 
 & 16  & \qsixteenGrid & \qsixteenRSG\\ \cline{2-4}
  & 25 & \qtwentyfiveGrid & \qtwentyfiveRSG  \\ \cline{2-4}
  & 36 & 11x11  & 25x25  \\ \hline
  
  \multirow{3}{1.8cm}{Quantum Approximate Optimization Alg (QAOA)} & 16  & \qsixteenGrid & \qsixteenRSG\\ \cline{2-4}
  & 25 & \qtwentyfiveGrid & \qtwentyfiveRSG  \\ \cline{2-4}
  & 36 & 11x11  & 25x25  \\ \hline
 
 \multirow{3}{1.8cm}{Ripple-Carry Adder (RCA)} 
 & 16  & \qsixteenGrid & \qsixteenRSG\\ \cline{2-4}
  & 25 & \qtwentyfiveGrid & \qtwentyfiveRSG  \\ \cline{2-4}
  & 36 & 11x11  & 25x25  \\ \hline
 \multirow{3}{1.8cm}{Bernstein Vazirani (BV)} 
 & 16  & \qsixteenGrid & \qsixteenRSG\\ \cline{2-4}
  & 25 & \qtwentyfiveGrid & \qtwentyfiveRSG  \\ \cline{2-4}
 & 100 & 19x19 & 43x43 \\ \hline
\end{tabular}
}
    \label{tab:benchmark}
\end{table}

\begin{table*}[h]
\centering
    \caption{The results of \frameworkname~and its relative performance to the baseline. }
\resizebox{\textwidth}{!}{
\renewcommand*{\arraystretch}{1.4}
\begin{tabular}{|p{2cm}|p{2cm}|p{2.5cm}|p{2cm}|p{2.4cm}|p{2.9cm}|p{2cm}|}  \hline
{Name-\#qubits} &  {Baseline Depth} & {Our Depth} & Improv. Factor & Baseline \#Fusions   & Our \#Fusions & Improv. Factor \\ \hline

QFT-16  & 787 & 83 & 9 & 201,472 & 8,167 & 25 \\\hline
QFT-25 & 1,518 & 162 & 9 & 669,438 & 26,921 & 25\\\hline
QFT-36 & 2,712 & 324 & 8 & 1,695,000 & 66,830 & 25 \\\hline

QAOA-16  & 595 & 29 & 20 & 152,320 & 2,578 & 59 \\\hline
QAOA-25  & 1,287 & 63 & 20 & 567,567 & 8,343 & 68\\\hline
QAOA-36  & 2,648 & 122 & 22 & 1,655,000 & 21,302 & 78\\\hline

RCA-16 & 734 & 46 & 16 & 187,904 & 4,568 & 41 \\\hline
RCA-25 & 1,273 & 65 & 20 & 561,393 & 8,915 & 63 \\\hline
RCA-36 & 1,934 & 85 & 23 & 1,208,750 & 14,115 & 86 \\\hline

BV-16 & 94 & 1 & 92 & 24,064 & 63 & 382\\\hline
BV-25 & 181 & 1 & 181 & 79,821 & 114 & 700\\\hline
BV-100 & 787 & 4 & 197 & 1,455,163 & 644 & 2,260\\\hline
\end{tabular}
    }

    \label{tab:evaluation}
\end{table*}

\subsection{Experiment Result}

We first show the results of  the baseline and \frameworkname\ for 3-qubit resource states in Table~\ref{tab:evaluation}, as the number of qubits increases for each benchmark. It can be seen that \frameworkname\ outperforms the baseline by orders of magnitude in both physical depth and the number of fusions. As the number of qubits increases, this improvement of performance stays stable or slightly increases.

Among the benchmarks, \frameworkname\ has the most significant improvement for BV. This is because the graph state of BV is acyclic and planar, making it easier the heuristic search in \frameworkname\ to find a compact layout for the graph and map it to only a few physical layers. In contrast, the graph states of the QFT, QAOA and RCA contain a lot of short-range and long-range cycles, which makes the structure more complex and increases the demand for routing. However, for these benchmarks, \frameworkname\ still achieves a reduction on the physical depth by a factor of 15 on average (geomean), and a reduction on the number of fusions by a factor of 47 on average (geomean).

\begin{figure}[h!]
        \centering
        \hspace{-15pt}\includegraphics[width=0.7\linewidth]{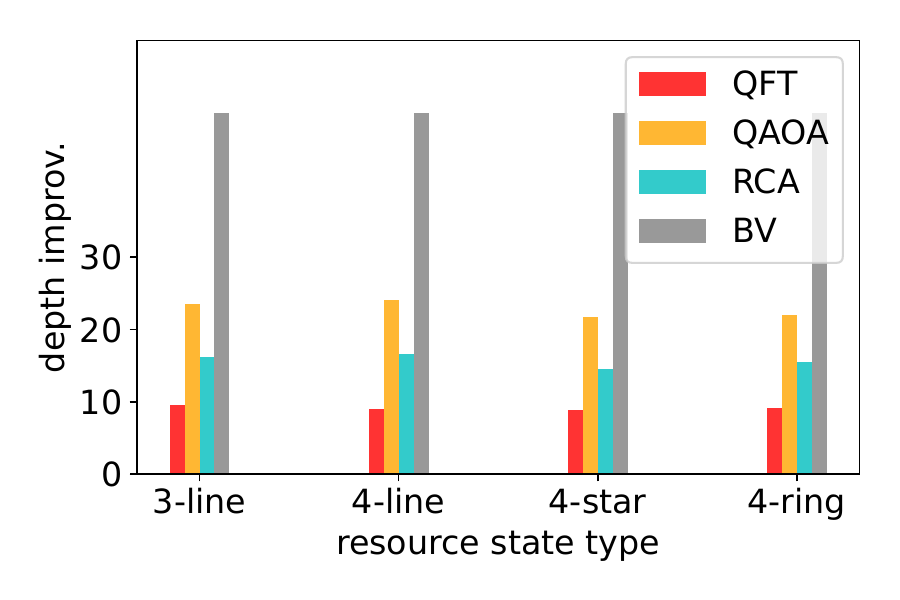} \\
        (a)\\
        \vspace{5pt}
        \hspace{-15pt}\includegraphics[width=0.7\linewidth]{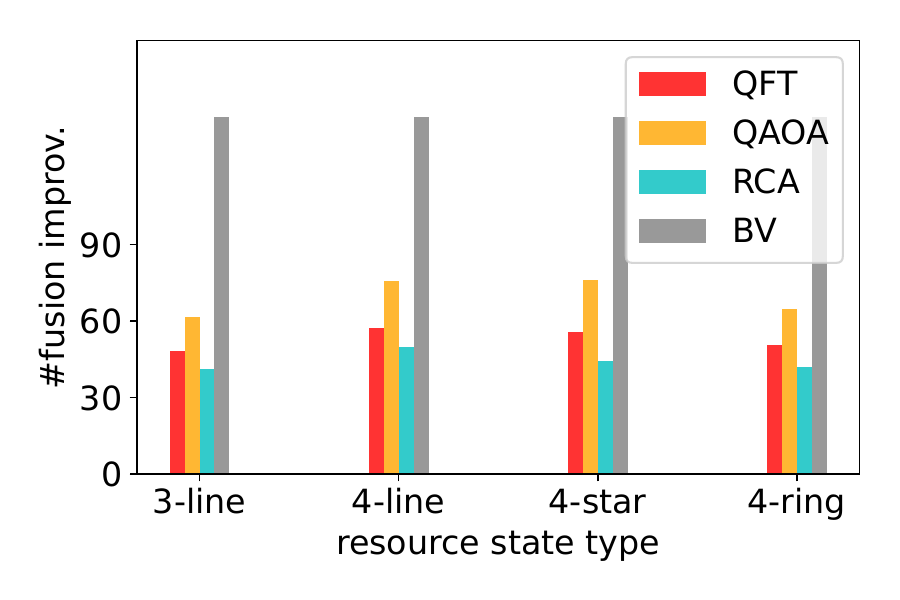}\\
        (b)
    
        \caption{\fix{Improvement factors of physical depth (a) and \#~fusions (b)  of 16-qubit QFT, QAOA, RAC and BV for different basic resource states.}
        }
        \label{fig:different_resource_states}
\end{figure}

\noindent\textbf{General resource states}
To achieve greater generality, our optimization modules are designed to be extendable to a wide range of resource states. 
This is made possible by accommodating flexible fusion strategies for synthesizing graph states and representing each strategy with a fusion graph.
Figure~\ref{fig:different_resource_states} shows the results when the resource states are 3-qubit line-shaped graph state, 4-qubit line-shaped, star-shaped and ring-shaped graphs states, respectively. The improvements in physical depth and the number of fusions are shown in (a) and (b), respectively. It can be seen that \frameworkname\ achieves similar levels of improvement in performance for these various resource states. 

\begin{figure}[h!]
        \centering
        \hspace{-15pt}\includegraphics[width=0.7\linewidth]{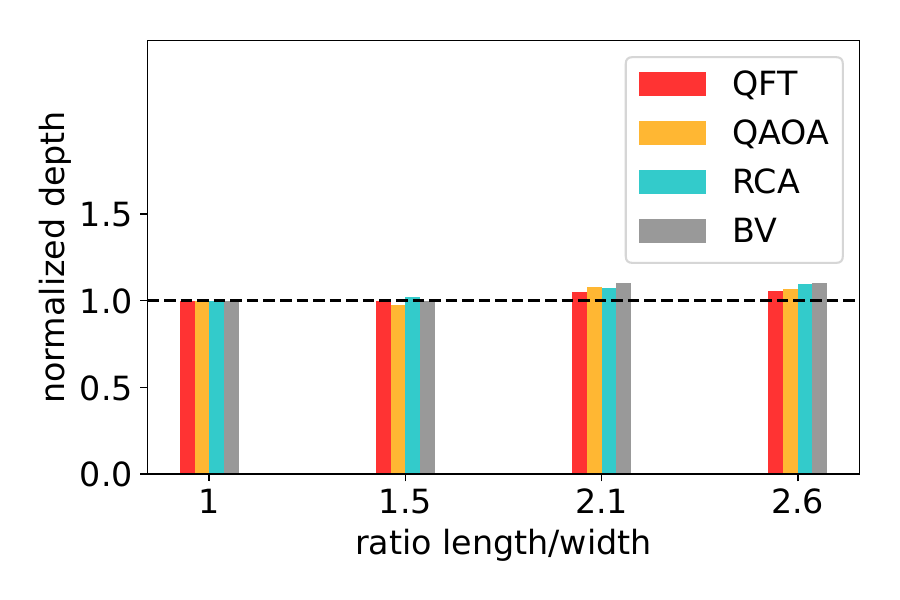} \\
        (a)\\
        \vspace{5pt}
        \hspace{-15pt}\includegraphics[width=0.7\linewidth]{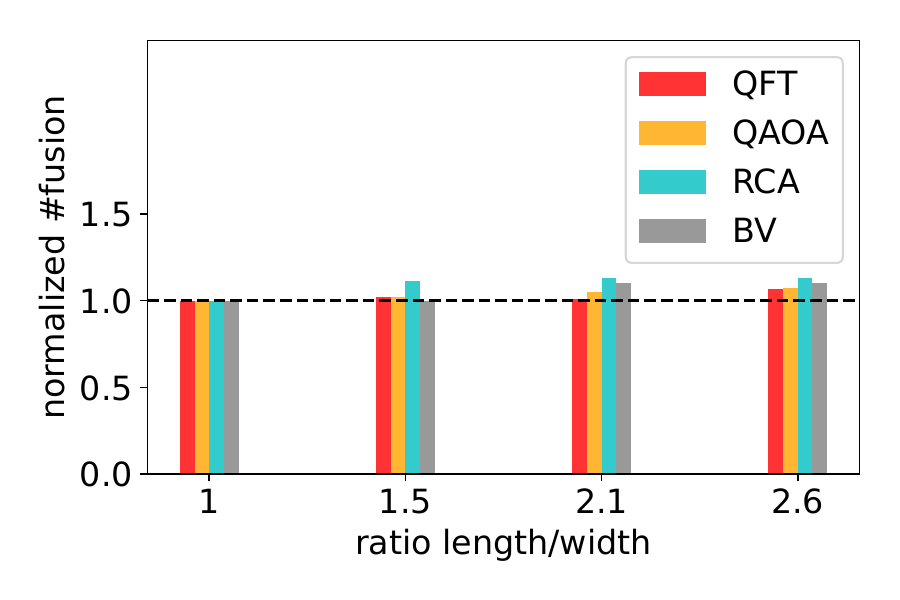}\\
        (b)
    
        \caption{Normalized physical depth (a) and \#~fusions (b) of 16-qubit QFT, QAOA, RCA and BV on different shapes of physical layers. Results on rectangular physical layers (ratio $>1$) are normalized by those on square physical layers (ratio $= 1$). 
        }
        \label{fig:different_layer_shapes}
\end{figure}

\begin{figure*}[t]
    \includegraphics[height=0.2\linewidth,width=0.7\linewidth]{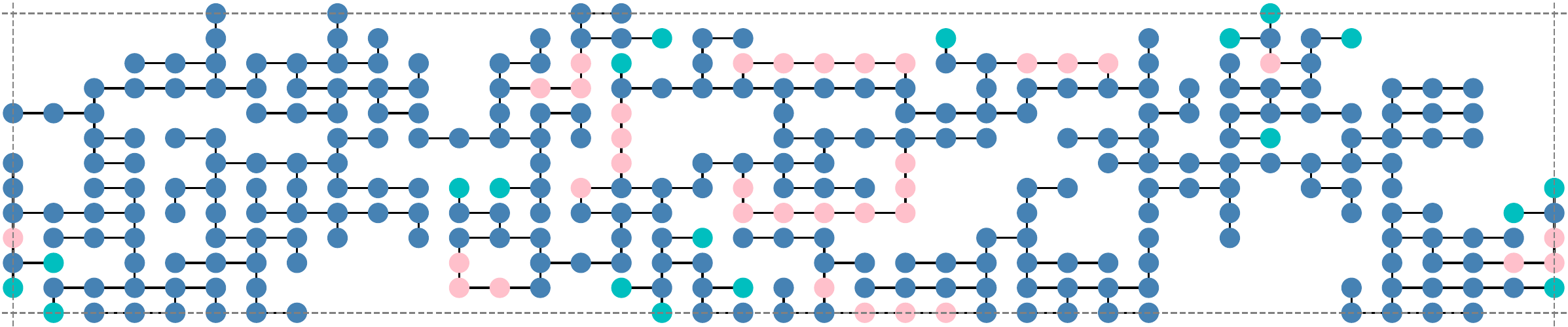}
        \\
        \caption{ Mapping of the fusion graph of a 16-qubit QFT program with a more global optimization by considering an extended physical layer consisting of 3 successive physical layers of size $13\times13$. Blue and green dots correspond to the nodes `$\otimes$' existing in the fusion graph, with the green ones being the incomplete nodes whose edges are not all mapped yet. Pink dots represent the auxiliary resource states used for routing.
        }
        \label{fig:extended_case_study}
\end{figure*}

\noindent \textbf{Flexible coupling structure} While the results in Table~\ref{tab:evaluation} are on physical layers of $N\times N$ square coupling structure, \frameworkname\ is applicable to more general coupling structures. Figure~\ref{fig:different_layer_shapes} shows the physical depth (\ref{fig:different_layer_shapes}(a)) and the number of fusions (\ref{fig:different_layer_shapes}(b)) when the ratio of each physical layer is 16x16  (ratio = 1), 20x13 (ratio = 1.5), 23x11  (ratio = 2.1) and 26x10 (ratio = 2.6) respectively. For comparison, the results are normalized by those on the square physical layer. It can be seen that \frameworkname\ achieves similar levels of performance on these various shapes of physical layers. Moreover, the area of physical layers is adjustable by regarding multiple consecutive physical layers as an extended physical layer, which can be used to accommodate more global structure of the graph. As a demonstration, Figure~\ref{fig:extended_case_study} shows a slice of mapping generated by \frameworkname~for a 16-qubit QFT program. The whole rectangular $13\times39$ extended physical layer is composed of 3 layers of $13\times 13$ grid. The first physical layer ranges from column 1 to column 13, the second ranges from column 14 to column 26, and the third from column 27 to column 39 respectively. In the figure, blue and green dots correspond to the nodes `$\otimes$' existing in the fusion graph, with the green ones representing the incomplete nodes, which means that some of edges connected to them have not been mapped onto the grid. Pink dots represent the auxiliary resource states used for routing. In principle, the optimization techniques in \frameworkname\ is also applicable when the coupling structure between RSGs are not orthogonal (e.g., triangular, hexagonal) since none of our optimization modules has explicit dependency on a specific coupling architecture.

\begin{figure}[h!]
        \centering
        \hspace{-15pt}\includegraphics[width=0.7\linewidth]{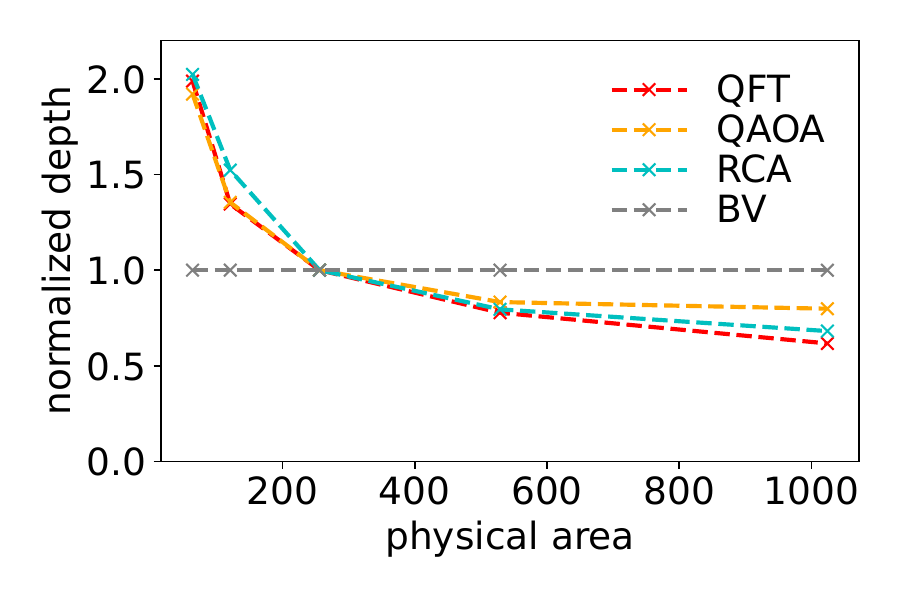} \\
        (a)\\
        \vspace{5pt}
        \hspace{-15pt}\includegraphics[width=0.7\linewidth]{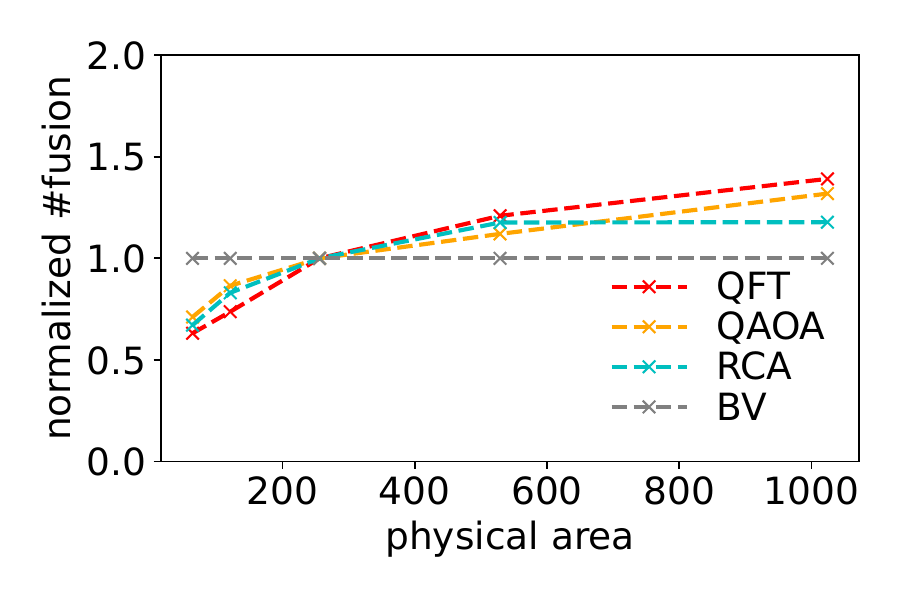}\\
        (b)
        \caption{\fix{Normalized physical depth (a) and \#~fusions (b) of 16-qubit QFT, QAOA, RAC and BV for different physical areas. Results on different physical areas are normalized by those on the same physical area as required by the baseline approach. }
        }
        \label{fig:depth_with_physical_area}
\end{figure}

\textbf{Effect of physical area}
In contrast to the fixed size of physical layers required by the baseline approach, \frameworkname\ does not impose strict constraints on the size of physical layers, allowing programs to be compiled onto hardware of different sizes. Specifically, the physical depth can be decreased by allowing longer routing paths on larger physical layers, with a cost of more fusions. Figure~\ref{fig:depth_with_physical_area} shows the physical depth and the number of fusions when the programs are compiled onto different sizes of physical layers. For comparison, the results are normalized by those on the physical area required by the baseline approach, which is 16x16 when the number of qubits is 16. It can be seen that when the physical area increases, the physical depth initially decreases rapidly, but eventually reaches a plateau when the physical area becomes sufficiently large. On the other hand, the number of fusions shows a reverse trend, increasing as the physical area increases. Note that if physical layers are larger than necessary, they can be truncated by restricting compilation to a portion of the physical layers.


%% file: 08_conclusion.tex
\vspace{0.3em}
\section{Conclusion}
In this work, we provide in-depth analysis and discussion of the unique challenges for  photonic one-way computing at both programming and hardware levels. We build the first end-to-end compilation framework for optimizing the mapping of any quantum programs onto photonic devices supported by our hardware abstract model. 
With this being said, many of the optimization problems here are NP-hard problems, and thus we believe there is still significant potential for fully exploring the entire optimization space. 
As the first work in this direction, we hope our work could attract more effort from the computer architecture and compiler community to explore the advantages of photonic quantum architectures and overcome the unique challenges. 

%% file: 09_acknoledgement.tex

\begin{acks}
We thank the anonymous reviewers for their constructive feedback and the cloud bank~\cite{cloud_credit}. This work is supported in part by Cisco Research, NSF 2048144, and Robert N.Noyce Trust.
\end{acks}

%% file: sample-sigconf-isca23.bbl
\begin{thebibliography}{10}

\bibitem{shor1999polynomial}
Peter~W Shor.
\newblock Polynomial-time algorithms for prime factorization and discrete
  logarithms on a quantum computer.
\newblock {\em SIAM review}, 41(2):303--332, 1999.

\bibitem{grover1996fast}
Lov~K Grover.
\newblock A fast quantum mechanical algorithm for database search.
\newblock In {\em Proceedings of the twenty-eighth annual ACM symposium on
  Theory of computing}, pages 212--219, 1996.

\bibitem{quantum_chemistry_1}
Yudong Cao, Jonathan Romero, Jonathan~P. Olson, Matthias Degroote, Peter~D.
  Johnson, M{\'{a}}ria Kieferov{\'{a}}, Ian~D. Kivlichan, Tim Menke, Borja
  Peropadre, Nicolas~P.D. Sawaya, Sukin Sim, Libor Veis, and Al{\'{a}}n
  Aspuru-Guzik.
\newblock {Quantum Chemistry in the Age of Quantum Computing}.
\newblock {\em Chemical Reviews}, 119:10856--10915, 2019.

\bibitem{quantum_chemistry_2}
Sam McArdle, Suguru Endo, Al{\'a}n Aspuru-Guzik, Simon~C Benjamin, and Xiao
  Yuan.
\newblock Quantum computational chemistry.
\newblock {\em Reviews of Modern Physics}, 92(1):015003, 2020.

\bibitem{devoret2013superconducting}
Michel~H Devoret and Robert~J Schoelkopf.
\newblock Superconducting circuits for quantum information: an outlook.
\newblock {\em Science}, 339(6124):1169--1174, 2013.

\bibitem{bruzewicz2019trapped}
Colin~D Bruzewicz, John Chiaverini, Robert McConnell, and Jeremy~M Sage.
\newblock Trapped-ion quantum computing: Progress and challenges.
\newblock {\em Applied Physics Reviews}, 6(2):021314, 2019.

\bibitem{saffman2016quantum}
Mark Saffman.
\newblock Quantum computing with atomic qubits and rydberg interactions:
  progress and challenges.
\newblock {\em Journal of Physics B: Atomic, Molecular and Optical Physics},
  49(20):202001, 2016.

\bibitem{photonics_review}
Pieter Kok, William~J Munro, Kae Nemoto, Timothy~C Ralph, Jonathan~P Dowling,
  and Gerard~J Milburn.
\newblock Linear optical quantum computing with photonic qubits.
\newblock {\em Reviews of modern physics}, 79(1):135, 2007.

\bibitem{takeda2019toward}
Shuntaro Takeda and Akira Furusawa.
\newblock Toward large-scale fault-tolerant universal photonic quantum
  computing.
\newblock {\em APL Photonics}, 4(6):060902, 2019.

\bibitem{pant2019percolation}
Mihir Pant, Don Towsley, Dirk Englund, and Saikat Guha.
\newblock Percolation thresholds for photonic quantum computing.
\newblock {\em Nature communications}, 10(1):1--11, 2019.

\bibitem{OBrien2009}
Jeremy~L. O'Brien, Akira Furusawa, and Jelena Vučković.
\newblock Photonic quantum technologies.
\newblock {\em Nature Photonics}, 3(12):687, 2009.

\bibitem{Bogdanov:17}
S.~Bogdanov, M.~Y. Shalaginov, A.~Boltasseva, and V.~M. Shalaev.
\newblock Material platforms for integrated quantum photonics.
\newblock {\em Opt. Mater. Express}, 7(1):111--132, Jan 2017.

\bibitem{larsen2021deterministic}
Mikkel~V Larsen, Xueshi Guo, Casper~R Breum, Jonas~S Neergaard-Nielsen, and
  Ulrik~L Andersen.
\newblock Deterministic multi-mode gates on a scalable photonic quantum
  computing platform.
\newblock {\em Nature Physics}, 17(9):1018--1023, 2021.

\bibitem{wang2020integrated}
Jianwei Wang, Fabio Sciarrino, Anthony Laing, and Mark~G Thompson.
\newblock Integrated photonic quantum technologies.
\newblock {\em Nature Photonics}, 14(5):273--284, 2020.

\bibitem{Madsen2022}
Lars~S. Madsen, Fabian Laudenbach, Mohsen~Falamarzi. Askarani, Fabien Rortais,
  Trevor Vincent, Jacob F.~F. Bulmer, Filippo~M. Miatto, Leonhard Neuhaus,
  Lukas~G. Helt, Matthew~J. Collins, Adriana~E. Lita, Thomas Gerrits, Sae~Woo
  Nam, Varun~D. Vaidya, Matteo Menotti, Ish Dhand, Zachary Vernon, Nicol{\'a}s
  Quesada, and Jonathan Lavoie.
\newblock Quantum computational advantage with a programmable photonic
  processor.
\newblock {\em Nature}, 606(7912):75--81, Jun 2022.

\bibitem{zhong2020quantum}
Han-Sen Zhong, Hui Wang, Yu-Hao Deng, Ming-Cheng Chen, Li-Chao Peng, Yi-Han
  Luo, Jian Qin, Dian Wu, Xing Ding, Yi~Hu, Peng Hu, Xiao-Yan Yang, Wei-Jun
  Zhang, Hao Li, Yuxuan Li, Xiao Jiang, Lin Gan, Guangwen Yang, Lixing You,
  Zhen Wang, Li~Li, Nai-Le Liu, Chao-Yang Lu, and Jian-Wei Pan.
\newblock Quantum computational advantage using photons.
\newblock {\em Science}, 370(6523):1460--1463, 2020.

\bibitem{PhysRevLett.127.180502}
Han-Sen Zhong, Yu-Hao Deng, Jian Qin, Hui Wang, Ming-Cheng Chen, Li-Chao Peng,
  Yi-Han Luo, Dian Wu, Si-Qiu Gong, Hao Su, Yi~Hu, Peng Hu, Xiao-Yan Yang,
  Wei-Jun Zhang, Hao Li, Yuxuan Li, Xiao Jiang, Lin Gan, Guangwen Yang, Lixing
  You, Zhen Wang, Li~Li, Nai-Le Liu, Jelmer~J. Renema, Chao-Yang Lu, and
  Jian-Wei Pan.
\newblock Phase-programmable gaussian boson sampling using stimulated squeezed
  light.
\newblock {\em Phys. Rev. Lett.}, 127:180502, Oct 2021.

\bibitem{bartolucci2021fusion}
Sara Bartolucci, Patrick Birchall, Hector Bombin, Hugo Cable, Chris Dawson,
  Mercedes Gimeno-Segovia, Eric Johnston, Konrad Kieling, Naomi Nickerson,
  Mihir Pant, Fernando Pastawski, Terry Rudolph, and Chris Sparrow.
\newblock Fusion-based quantum computation.
\newblock {\em arXiv preprint arXiv:2101.09310}, 2021.

\bibitem{bombin2021interleaving}
Hector Bombin, Isaac~H Kim, Daniel Litinski, Naomi Nickerson, Mihir Pant,
  Fernando Pastawski, Sam Roberts, and Terry Rudolph.
\newblock Interleaving: Modular architectures for fault-tolerant photonic
  quantum computing.
\newblock {\em arXiv preprint arXiv:2103.08612}, 2021.

\bibitem{bartolucci2021switch}
Sara Bartolucci, Patrick Birchall, Damien Bonneau, Hugo Cable, Mercedes
  Gimeno-Segovia, Konrad Kieling, Naomi Nickerson, Terry Rudolph, and Chris
  Sparrow.
\newblock Switch networks for photonic fusion-based quantum computing.
\newblock {\em arXiv preprint arXiv:2109.13760}, 2021.

\bibitem{Qiskit}
MD~SAJID ANIS, H{\'e}ctor Abraham, AduOffei, Rochisha Agarwal, Gabriele
  Agliardi, Merav Aharoni, Ismail~Yunus Akhalwaya, Gadi Aleksandrowicz, Thomas
  Alexander, Matthew Amy, Sashwat Anagolum, Eli Arbel, Abraham Asfaw, Anish
  Athalye, Artur Avkhadiev, Carlos Azaustre, PRATHAMESH BHOLE, Abhik Banerjee,
  Santanu Banerjee, Will Bang, Aman Bansal, Panagiotis Barkoutsos, Ashish
  Barnawal, George Barron, George~S. Barron, Luciano Bello, Yael Ben-Haim,
  M.~Chandler Bennett, Daniel Bevenius, Dhruv Bhatnagar, Arjun Bhobe, Paolo
  Bianchini, Lev~S. Bishop, Carsten Blank, Sorin Bolos, Soham Bopardikar,
  Samuel Bosch, Sebastian Brandhofer, Brandon, Sergey Bravyi, Nick Bronn,
  Bryce-Fuller, David Bucher, Artemiy Burov, Fran Cabrera, Padraic Calpin,
  Lauren Capelluto, Jorge Carballo, Gin{\'e}s Carrascal, Adam Carriker, Ivan
  Carvalho, Adrian Chen, Chun-Fu Chen, Edward Chen, Jielun~(Chris) Chen,
  Richard Chen, Franck Chevallier, Kartik Chinda, Rathish Cholarajan, Jerry~M.
  Chow, Spencer Churchill, CisterMoke, Christian Claus, Christian Clauss, Caleb
  Clothier, Romilly Cocking, Ryan Cocuzzo, Jordan Connor, Filipe Correa,
  Abigail~J. Cross, Andrew~W. Cross, Simon Cross, Juan Cruz-Benito, Chris
  Culver, Antonio~D. C{\'o}rcoles-Gonzales, Navaneeth D, Sean Dague, Tareq~El
  Dandachi, Animesh~N Dangwal, Jonathan Daniel, Marcus Daniels, Matthieu
  Dartiailh, Abd{\'o}n~Rodr{\'\i}guez Davila, Faisal Debouni, Anton Dekusar,
  Amol Deshmukh, Mohit Deshpande, Delton Ding, Jun Doi, Eli~M. Dow, Eric
  Drechsler, Eugene Dumitrescu, Karel Dumon, Ivan Duran, Kareem EL-Safty, Eric
  Eastman, Grant Eberle, Amir Ebrahimi, Pieter Eendebak, Daniel Egger, ElePT,
  Emilio, Alberto Espiricueta, Mark Everitt, Davide Facoetti, Farida,
  Paco~Mart{\'\i}n Fern{\'a}ndez, Samuele Ferracin, Davide Ferrari,
  Axel~Hern{\'a}ndez Ferrera, Romain Fouilland, Albert Frisch, Andreas Fuhrer,
  Bryce Fuller, MELVIN GEORGE, Julien Gacon, Borja~Godoy Gago, Claudio
  Gambella, Jay~M. Gambetta, Adhisha Gammanpila, Luis Garcia, Tanya Garg,
  Shelly Garion, James~R. Garrison, Tim Gates, Leron Gil, Austin Gilliam,
  Aditya Giridharan, Juan Gomez-Mosquera, Gonzalo, Salvador de~la
  Puente~Gonz{\'a}lez, Jesse Gorzinski, Ian Gould, Donny Greenberg, Dmitry
  Grinko, Wen Guan, Dani Guijo, John~A. Gunnels, Harshit Gupta, Naman Gupta,
  Jakob~M. G{\"u}nther, Mikael Haglund, Isabel Haide, Ikko Hamamura, Omar~Costa
  Hamido, Frank Harkins, Kevin Hartman, Areeq Hasan, Vojtech Havlicek, Joe
  Hellmers, {\L}ukasz Herok, Stefan Hillmich, Hiroshi Horii, Connor Howington,
  Shaohan Hu, Wei Hu, Junye Huang, Rolf Huisman, Haruki Imai, Takashi Imamichi,
  Kazuaki Ishizaki, Ishwor, Raban Iten, Toshinari Itoko, Alexander Ivrii, Ali
  Javadi, Ali Javadi-Abhari, Wahaj Javed, Qian Jianhua, Madhav Jivrajani, Kiran
  Johns, Scott Johnstun, Jonathan-Shoemaker, JosDenmark, JoshDumo, John Judge,
  Tal Kachmann, Akshay Kale, Naoki Kanazawa, Jessica Kane, Kang-Bae, Annanay
  Kapila, Anton Karazeev, Paul Kassebaum, Josh Kelso, Scott Kelso, Vismai
  Khanderao, Spencer King, Yuri Kobayashi, Kovi11Day, Arseny Kovyrshin, Rajiv
  Krishnakumar, Vivek Krishnan, Kevin Krsulich, Prasad Kumkar, Gawel Kus, Ryan
  LaRose, Enrique Lacal, Rapha{\"e}l Lambert, Haggai Landa, John Lapeyre, Joe
  Latone, Scott Lawrence, Christina Lee, Gushu Li, Jake Lishman, Dennis Liu,
  Peng Liu, Abhishek~K M, Liam Madden, Yunho Maeng, Saurav Maheshkar, Kahan
  Majmudar, Aleksei Malyshev, Mohamed~El Mandouh, Joshua Manela, Manjula, Jakub
  Marecek, Manoel Marques, Kunal Marwaha, Dmitri Maslov, Pawe{\l} Maszota,
  Dolph Mathews, Atsushi Matsuo, Farai Mazhandu, Doug McClure, Maureen
  McElaney, Cameron McGarry, David McKay, Dan McPherson, Srujan Meesala, Dekel
  Meirom, Corey Mendell, Thomas Metcalfe, Martin Mevissen, Andrew Meyer,
  Antonio Mezzacapo, Rohit Midha, Daniel Miller, Zlatko Minev, Abby Mitchell,
  Nikolaj Moll, Alejandro Montanez, Gabriel Monteiro, Michael~Duane Mooring,
  Renier Morales, Niall Moran, David Morcuende, Seif Mostafa, Mario Motta,
  Romain Moyard, Prakash Murali, Jan M{\"u}ggenburg, Tristan NEMOZ, David
  Nadlinger, Ken Nakanishi, Giacomo Nannicini, Paul Nation, Edwin Navarro,
  Yehuda Naveh, Scott~Wyman Neagle, Patrick Neuweiler, Aziz Ngoueya, Johan
  Nicander, Nick-Singstock, Pradeep Niroula, Hassi Norlen, NuoWenLei, Lee~James
  O'Riordan, Oluwatobi Ogunbayo, Pauline Ollitrault, Tamiya Onodera, Raul
  Otaolea, Steven Oud, Dan Padilha, Hanhee Paik, Soham Pal, Yuchen Pang, Ashish
  Panigrahi, Vincent~R. Pascuzzi, Simone Perriello, Eric Peterson, Anna Phan,
  Kuba Pilch, Francesco Piro, Marco Pistoia, Christophe Piveteau, Julia Plewa,
  Pierre Pocreau, Alejandro Pozas-Kerstjens, Rafa{\l} Pracht, Milos Prokop,
  Viktor Prutyanov, Sumit Puri, Daniel Puzzuoli, Jes{\'u}s P{\'e}rez, Quant02,
  Quintiii, Rafey~Iqbal Rahman, Arun Raja, Roshan Rajeev, Isha Rajput, Nipun
  Ramagiri, Anirudh Rao, Rudy Raymond, Oliver Reardon-Smith, Rafael
  Mart{\'\i}n-Cuevas Redondo, Max Reuter, Julia Rice, Matt Riedemann, Rietesh,
  Drew Risinger, Marcello~La Rocca, Diego~M. Rodr{\'\i}guez, RohithKarur, Ben
  Rosand, Max Rossmannek, Mingi Ryu, Tharrmashastha SAPV, Nahum Rosa~Cruz Sa,
  Arijit Saha, Abdullah Ash-Saki, Sankalp Sanand, Martin Sandberg, Hirmay
  Sandesara, Ritvik Sapra, Hayk Sargsyan, Aniruddha Sarkar, Ninad Sathaye,
  Bruno Schmitt, Chris Schnabel, Zachary Schoenfeld, Travis~L. Scholten, Eddie
  Schoute, Mark Schulterbrandt, Joachim Schwarm, James Seaward, Sergi,
  Ismael~Faro Sertage, Kanav Setia, Freya Shah, Nathan Shammah, Rohan Sharma,
  Yunong Shi, Jonathan Shoemaker, Adenilton Silva, Andrea Simonetto, Deeksha
  Singh, Divyanshu Singh, Parmeet Singh, Phattharaporn Singkanipa, Yukio
  Siraichi, Siri, Jes{\'u}s Sistos, Iskandar Sitdikov, Seyon Sivarajah,
  Magnus~Berg Sletfjerding, John~A. Smolin, Mathias Soeken, Igor~Olegovich
  Sokolov, Igor Sokolov, Vicente~P. Soloviev, SooluThomas, Starfish, Dominik
  Steenken, Matt Stypulkoski, Adrien Suau, Shaojun Sun, Kevin~J. Sung, Makoto
  Suwama, Oskar S{\l}owik, Hitomi Takahashi, Tanvesh Takawale, Ivano
  Tavernelli, Charles Taylor, Pete Taylour, Soolu Thomas, Kevin Tian, Mathieu
  Tillet, Maddy Tod, Miroslav Tomasik, Caroline Tornow, Enrique de~la Torre,
  Juan Luis~S{\'a}nchez Toural, Kenso Trabing, Matthew Treinish, Dimitar
  Trenev, TrishaPe, Felix Truger, Georgios Tsilimigkounakis, Davindra Tulsi,
  Wes Turner, Yotam Vaknin, Carmen~Recio Valcarce, Francois Varchon, Adish
  Vartak, Almudena~Carrera Vazquez, Prajjwal Vijaywargiya, Victor Villar,
  Bhargav Vishnu, Desiree Vogt-Lee, Christophe Vuillot, James Weaver, Johannes
  Weidenfeller, Rafal Wieczorek, Jonathan~A. Wildstrom, Jessica Wilson, Erick
  Winston, WinterSoldier, Jack~J. Woehr, Stefan Woerner, Ryan Woo,
  Christopher~J. Wood, Ryan Wood, Steve Wood, James Wootton, Matt Wright, Lucy
  Xing, Jintao YU, Bo~Yang, Unchun Yang, Daniyar Yeralin, Ryota Yonekura, David
  Yonge-Mallo, Ryuhei Yoshida, Richard Young, Jessie Yu, Lebin Yu, Christopher
  Zachow, Laura Zdanski, Helena Zhang, Iulia Zidaru, and Christa Zoufal.
\newblock Qiskit: An open-source framework for quantum computing, 2021.

\bibitem{sivarajah2020t}
Seyon Sivarajah, Silas Dilkes, Alexander Cowtan, Will Simmons, Alec Edgington,
  and Ross Duncan.
\newblock t| ket>: a retargetable compiler for nisq devices.
\newblock {\em Quantum Science and Technology}, 6(1):014003, 2020.

\bibitem{mbqc2009}
Robert Raussendorf.
\newblock Measurement-based quantum computation with cluster states.
\newblock {\em International Journal of Quantum Information}, 7(06):1053--1203,
  2009.

\bibitem{Xanadu}
Nathan Killoran, Josh Izaac, Nicol{\'a}s Quesada, Ville Bergholm, Matthew Amy,
  and Christian Weedbrook.
\newblock Strawberry fields: A software platform for photonic quantum
  computing.
\newblock {\em Quantum}, 3:129, 2019.

\bibitem{hamilton2017gaussian}
Craig~S Hamilton, Regina Kruse, Linda Sansoni, Sonja Barkhofen, Christine
  Silberhorn, and Igor Jex.
\newblock Gaussian boson sampling.
\newblock {\em Physical review letters}, 119(17):170501, 2017.

\bibitem{raussendorf2003measurement}
Robert Raussendorf, Daniel~E Browne, and Hans~J Briegel.
\newblock Measurement-based quantum computation on cluster states.
\newblock {\em Physical review A}, 68(2):022312, 2003.

\bibitem{briegel2009measurement}
Hans~J Briegel, David~E Browne, Wolfgang D{\"u}r, Robert Raussendorf, and
  Maarten Van~den Nest.
\newblock Measurement-based quantum computation.
\newblock {\em Nature Physics}, 5(1):19--26, 2009.

\bibitem{van2006universal}
Maarten Van~den Nest, Akimasa Miyake, Wolfgang D{\"u}r, and Hans~J Briegel.
\newblock Universal resources for measurement-based quantum computation.
\newblock {\em Physical review letters}, 97(15):150504, 2006.

\bibitem{GHZcluster}
Mercedes Gimeno-Segovia, Pete Shadbolt, Dan~E Browne, and Terry Rudolph.
\newblock From three-photon ghz states to universal ballistic quantum
  computation.
\newblock 2015.

\bibitem{con-rev-255}
Hans~J Briegel and Robert Raussendorf.
\newblock A one-way quantum computer.
\newblock {\em Physical Review Letters}, 86(22), 2003.

\bibitem{con-rev-256}
Michael~A Nielsen.
\newblock Optical quantum computation using cluster states.
\newblock {\em Physical review letters}, 93(4):040503, 2004.

\bibitem{con-rev-209}
S{\'e}bastien Tanzilli, Anthony Martin, Florian Kaiser, Marc~P De~Micheli,
  Olivier Alibart, and Daniel~B Ostrowsky.
\newblock On the genesis and evolution of integrated quantum optics.
\newblock {\em Laser \& Photonics Reviews}, 6(1):115--143, 2012.

\bibitem{mbqc-grover}
Philip Walther, Kevin~J Resch, Terry Rudolph, Emmanuel Schenck, Harald
  Weinfurter, Vlatko Vedral, Markus Aspelmeyer, and Anton Zeilinger.
\newblock Experimental one-way quantum computing.
\newblock {\em Nature}, 434(7030):169--176, 2005.

\bibitem{mbqc-dj}
Giuseppe Vallone, Gaia Donati, Natalia Bruno, Andrea Chiuri, and Paolo
  Mataloni.
\newblock Experimental realization of the deutsch-jozsa algorithm with a
  six-qubit cluster state.
\newblock {\em Physical Review A}, 81(5):050302, 2010.

\bibitem{mbqc-simon}
Mark~S Tame, Bryn~A Bell, Carlo Di~Franco, William~J Wadsworth, and John~G
  Rarity.
\newblock Experimental realization of a one-way quantum computer algorithm
  solving simon’s problem.
\newblock {\em Physical Review Letters}, 113(20):200501, 2014.

\bibitem{con-rev-102}
S{\'e}bastien Tanzilli, Hugues De~Riedmatten, Wolfgang Tittel, Hugo Zbinden,
  Pascal Baldi, Marc De~Micheli, Daniel~Barry Ostrowsky, and Nicolas Gisin.
\newblock Highly efficient photon-pair source using periodically poled lithium
  niobate waveguide.
\newblock {\em Electronics Letters}, 37(1):26--28, 2001.

\bibitem{con-rev-103}
Kaoru Sanaka, Karin Kawahara, and Takahiro Kuga.
\newblock New high-efficiency source of photon pairs for engineering quantum
  entanglement.
\newblock {\em Physical Review Letters}, 86(24):5620, 2001.

\bibitem{con-rev-104}
Konrad Banaszek, Alfred~B U’Ren, and Ian~A Walmsley.
\newblock Generation of correlated photons in controlled spatial modes by
  downconversion in nonlinear waveguides.
\newblock {\em Optics letters}, 26(17):1367--1369, 2001.

\bibitem{con-rev-217}
Simone Ferrari, Carsten Schuck, and Wolfram Pernice.
\newblock Waveguide-integrated superconducting nanowire single-photon
  detectors.
\newblock {\em Nanophotonics}, 7(11):1725--1758, 2018.

\bibitem{con-rev-204}
Jianwei Wang, Stefano Paesani, Yunhong Ding, Raffaele Santagati, Paul
  Skrzypczyk, Alexia Salavrakos, Jordi Tura, Remigiusz Augusiak, Laura
  Man{\v{c}}inska, Davide Bacco, et~al.
\newblock Multidimensional quantum entanglement with large-scale integrated
  optics.
\newblock {\em Science}, 360(6386):285--291, 2018.

\bibitem{ferreira2022deterministic}
Vinicius~S Ferreira, Gihwan Kim, Andreas Butler, Hannes Pichler, and Oskar
  Painter.
\newblock Deterministic generation of multidimensional photonic cluster states
  with a single quantum emitter.
\newblock {\em arXiv preprint arXiv:2206.10076}, 2022.

\bibitem{con-rev-247}
Peter~J Shadbolt, Maria~R Verde, Alberto Peruzzo, Alberto Politi, Anthony
  Laing, Mirko Lobino, Jonathan~CF Matthews, Mark~G Thompson, and Jeremy~L
  O'Brien.
\newblock Generating, manipulating and measuring entanglement and mixture with
  a reconfigurable photonic circuit.
\newblock {\em Nature Photonics}, 6(1):45--49, 2012.

\bibitem{con-rev-252}
Jacques Carolan, Christopher Harrold, Chris Sparrow, Enrique
  Mart{\'\i}n-L{\'o}pez, Nicholas~J Russell, Joshua~W Silverstone, Peter~J
  Shadbolt, Nobuyuki Matsuda, Manabu Oguma, Mikitaka Itoh, et~al.
\newblock Universal linear optics.
\newblock {\em Science}, 349(6249):711--716, 2015.

\bibitem{psiquantum-fbqc}
Sara Bartolucci, Patrick Birchall, Hector Bombin, Hugo Cable, Chris Dawson,
  Mercedes Gimeno-Segovia, Eric Johnston, Konrad Kieling, Naomi Nickerson,
  Mihir Pant, Fernando Pastawski, Terry Rudolph, and Chris Sparrow.
\newblock Fusion-based quantum computation.
\newblock {\em arXiv preprint 2101.09310}, 2021.

\bibitem{con-rev-316}
Fabian Ewert and Peter van Loock.
\newblock 3/4-efficient bell measurement with passive linear optics and
  unentangled ancillae.
\newblock {\em Physical review letters}, 113(14):140403, 2014.

\bibitem{translation}
Anne Broadbent and Elham Kashefi.
\newblock Parallelizing quantum circuits.
\newblock {\em Theoretical computer science}, 410(26):2489--2510, 2009.

\bibitem{kissinger2019pyzx}
Aleks Kissinger and John van~de Wetering.
\newblock Pyzx: Large scale automated diagrammatic reasoning.
\newblock {\em arXiv preprint arXiv:1904.04735}, 2019.

\bibitem{sabre}
Gushu Li, Yufei Ding, and Yuan Xie.
\newblock Tackling the qubit mapping problem for nisq-era quantum devices.
\newblock In {\em Proceedings of the Twenty-Fourth International Conference on
  Architectural Support for Programming Languages and Operating Systems}, pages
  1001--1014, 2019.

\bibitem{networkx}
Aric Hagberg, Pieter Swart, and Daniel S~Chult.
\newblock Exploring network structure, dynamics, and function using networkx.
\newblock Technical report, Los Alamos National Lab.(LANL), Los Alamos, NM
  (United States), 2008.

\bibitem{compactness_NPhard}
Maurizio Patrignani.
\newblock On the complexity of orthogonal compaction.
\newblock {\em Computational Geometry}, 19(1):47--67, 2001.

\bibitem{RCA}
Steven~A Cuccaro, Thomas~G Draper, Samuel~A Kutin, and David~Petrie Moulton.
\newblock A new quantum ripple-carry addition circuit.
\newblock {\em arXiv preprint quant-ph/0410184}, 2004.

\bibitem{cloud_credit}
Michael Norman, Vince Kellen, Shava Smallen, Brian DeMeulle, Shawn Strande,
  Ed~Lazowska, Naomi Alterman, Rob Fatland, Sarah Stone, Amanda Tan, et~al.
\newblock Cloudbank: Managed services to simplify cloud access for computer
  science research and education.
\newblock In {\em Practice and Experience in Advanced Research Computing},
  pages 1--4. 2021.

\end{thebibliography}
